\documentclass[aps,twocolumn,prd,showpacs,showkeys,preprintnumbers,superscriptaddress,nofootinbib,floatfix,longbibliography]{revtex4-2}[prd]

\usepackage{orcidlink}
\usepackage{lipsum}
\usepackage{amsmath}
\usepackage{amsfonts}
\usepackage{amssymb}
\usepackage{mathrsfs}
\usepackage{graphicx}
\usepackage{color}
\usepackage[dvipsnames]{xcolor}
\usepackage{bm}
\usepackage{blindtext}
\usepackage{wasysym}
\usepackage{hyperref}
\hypersetup{colorlinks=true,allcolors=blue}

\usepackage[normalem]{ulem}
\usepackage{fontawesome} 
\usepackage[all]{hypcap}
\usepackage{lineno}

\newcommand{\ldm}{\ensuremath{{\Delta m_{31}^2}}}         
\newcommand{\sdm}{\ensuremath{{\Delta m_{21}^2}}}

\begin{document}
	
	\title{$\delta_{CP}$-free constraints on NSI parameters $\varepsilon_{e\mu}$ and $\varepsilon_{e\tau}$ using high-purity $\nu_\mu$ CC  events at IceCube DeepCore}
	
	\author{J Krishnamoorthi\,\orcidlink{0009-0006-1352-2248}}
	\email{krishnamoorthi.j@iopb.res.in}
	\affiliation{Institute of Physics, Sachivalaya Marg, Sainik School Post, Bhubaneswar 751005, India}
	\affiliation{Department of Physics, Aligarh Muslim University, Aligarh 202002, India}
	
	\author{Anil Kumar\,\orcidlink{0000-0002-8367-8401}}
	\email{anil.k@iopb.res.in}
	\affiliation{Institute of Physics, Sachivalaya Marg, Sainik School Post, Bhubaneswar 751005, India}
	
	\author{Sanjib Kumar Agarwalla\,\orcidlink{0000-0002-9714-8866}}
	\email{sanjib@iopb.res.in}
	\affiliation{Institute of Physics, Sachivalaya Marg, Sainik School Post, Bhubaneswar 751005, India}
	\affiliation{Homi Bhabha National Institute, Training School Complex, Anushakti Nagar, Mumbai 400094, India}
	
	\preprint{IOP/BBSR/2026-05}
	
	\date{\today}

\begin{abstract}	
	Atmospheric neutrinos provide a unique avenue to probe theories beyond the Standard Model (BSM) over a wide range of energies and path lengths. The theory of nonstandard interactions (NSI) of neutrinos is one of the important BSM scenarios, which can modify flavor oscillations of atmospheric neutrinos traveling through the Earth. In this work, we use a high-purity $\nu_{\mu}$ charged-current (CC) sample of atmospheric neutrinos from IceCube DeepCore with a livetime of 7.5 years to search for the NSI parameters $\varepsilon_{e\mu}$, $\varepsilon_{e\tau}$, and $\varepsilon_{ee}-\varepsilon_{\mu\mu}$. The $\nu_{\mu}$ CC events mainly come from the $\nu_{\mu}$ survival channel having no significant dependence on $\delta_{CP}$. Therefore, the constraints on $\varepsilon_{e\mu}$ and $\varepsilon_{e\tau}$ obtained using this $\nu_{\mu}$ CC sample are expected to be free from the $\delta_{CP}$-degeneracy. The data sample is found to be in agreement with the standard neutrino interactions. Therefore, we place bounds on these NSI parameters that are consistent with and comparable to existing experimental constraints. These $\delta_{CP}$-free constraints from IceCube DeepCore are complementary to those from the long-baseline neutrino oscillation experiments, where the appearance channel depends on $\delta_{CP}$.
\end{abstract}

\maketitle

\section{Introduction and motivation}\label{sec:introduction}

Neutrino oscillations~\cite{Super-Kamiokande:1998kpq,SNO:2001kpb,SNO:2002tuh} provide a strong evidence for the theories beyond the Standard Model (BSM) of particle physics by showing that neutrinos have nonzero masses, and flavor eigenstates ($\nu_e$, $\nu_{\mu}$, and $\nu_{\tau}$) are superpositions of mass eigenstates ($\nu_1$, $\nu_{2}$, and $\nu_{3}$). The relative phases of these mass eigenstates evolve during propagation according to quantum time evolution, which leads to neutrino flavor oscillations. The mixing between the flavor and mass eigenstates is described by the Pontecorvo-Maki-Nakagawa-Sakata (PMNS)~\cite{Maki:1962mu,Pontecorvo:1967fh} matrix, which is parameterized in terms of the three mixing angles ($\theta_{12}$, $\theta_{13}$, and $\theta_{23}$) and one Dirac-{\it CP} phase ($\delta_{CP}$). Neutrino oscillation probabilities depend upon the parameters of the PMNS matrix as well as the mass-squared differences ($\Delta m_{31}^2$ and $\Delta m_{21}^2$). During the last few decades, the neutrino oscillation parameters have been measured with good precision using the data from various neutrino experiments~\cite{ParticleDataGroup:2024cfk,Esteban:2024eli,Capozzi:2025ovi}. However, we still have some important unknowns in the field of neutrino oscillations, which include the measurement of $\delta_{CP}$, determination of the octant of $\theta_{23}$, and the neutrino mass ordering (i.e., the sign of $\Delta m_{31}^2$).

Even though we are in the precision era of neutrino oscillations, the mechanism that provides the neutrino mass is still unknown. The neutrino mass generation mechanism is expected to be associated with new physics, which can also lead to new interactions of neutrinos with matter. These new interactions are commonly referred to as nonstandard interactions (NSI) of neutrinos. By considering the Standard Model (SM) as an effective field theory~\cite{Weinberg:1967tq}, NSI can be parameterized by dimension-six four-fermion operators~\cite{Gavela:2008ra, Bischer:2019ttk}, which can be sourced in various BSM scenarios. NSI can be categorized into two types: neutral-current (NC) NSI and charged-current (CC) NSI. The NC-NSI can affect the neutrino propagation through matter, while the CC-NSI can modify the production and detection processes of neutrinos. In this work, we focus only on the NC-NSI, which have been studied extensively in literature~\cite{Valle:1987gv,Roulet:1991sm, Guzzo:1991hi,Guzzo:2000kx,Huber:2001zw,Gonzalez-Garcia:2004pka,Kopp:2008,Biggio:2009nt,Escrihuela:2011cf,Gonzalez-Garcia:2011vlg,Agarwalla:2012wf,Ohlsson:2012kf,Esmaili:2013fva,Gonzalez-Garcia:2013usa,MINOS:2013hmj,Chatterjee:2014gxa,Agarwalla:2014bsa,Mocioiu:2014gua,Miranda:2015dra,Agarwalla:2015cta,Choubey:2015xha,Agarwalla:2016fkh,Salvado:2016uqu,Farzan:2017xzy,Coloma:2017egw,IceCube:2017zcu,Esteban:2018ppq,Borexino:2019mhy,Bhupal:2019qno,Khatun:2020,Kumar:2021lrn,Agarwalla:2021zfr,IceCubeCollaboration:2021euf,IceCube:2022pbe,Krishnamoorthi:2025,Coloma:2023ixt,NOvA:2024lti,IceCubeCollaboration:2021euf,KM3NeT:2024pte,ANTARES:2021crm, IceCube:2022ubv,Super-Kamiokande:2011dam,Jana:2024lfm,Krishnamoorthi:2025efw}.

The extended Lagrangian for the NC-NSI can be written as
\begin{equation}
	\mathcal{L}_{\rm NSI}^{\rm NC} = -2\sqrt{2} G_F \varepsilon_{\alpha\beta}^{fC} (\bar{\nu}_\alpha \gamma^\mu P_L \nu_\beta)(\bar{f} \gamma_\mu P_C f)\,,
\end{equation}
where $G_F$ is the Fermi coupling constant and $\varepsilon_{\alpha\beta}^{fC}$ represent the NSI coupling strength with $\alpha$ and $\beta$ being the neutrino flavor indices. Here, $f$ represents the matter fermions (electrons, up quark, and down quark).  $P_C$ is the projection operator for the left-handed ($C=L$) or right-handed ($C=R$) chirality. The NSI parameters can be flavor-conserving (nonuniversal), when $\alpha=\beta$, or flavor-violating when $\alpha \neq \beta$. Since we consider the new interaction of neutrinos with the matter fermions in the Earth, the NSI coupling strength is the sum of the left-handed and right-handed chirality fermions, i.e.,  $\varepsilon_{\alpha\beta}^{f}=\varepsilon_{\alpha\beta}^{fL}+\varepsilon_{\alpha\beta}^{fR}$. When neutrinos propagate through Earth matter, they experience coherent forward scattering with all matter fermions, which can be described by an effective NSI parameter defined as
\begin{equation}
	\varepsilon_{\alpha\beta} = \sum_{f=e,u,d} \frac{N_f}{N_e} \varepsilon_{\alpha\beta}^f\,,
\end{equation}
where $N_f$ and $N_e$ are the number densities of the fermions $f$ and electrons, respectively, in the Earth. The NSI parameter $\varepsilon_{\alpha\beta}$ acts as an effective matter potential that can affect the oscillation probabilities of neutrinos propagating through the Earth. 

In this work, we present constraints on the NSI parameters $\varepsilon_{e\mu}$, $\varepsilon_{e\tau}$, and $\varepsilon_{ee}-\varepsilon_{\mu\mu}$ considering one parameter at a time using the publicly available IceCube DeepCore atmospheric neutrino data~\cite{DVN_B4RITM_2025} with about 7.5 years of livetime. This data sample consists of high-quality atmospheric neutrino events in the energy range of 6.3 GeV to 158.5 GeV, where only those Cherenkov photons have been selected that directly reach the detectors without experiencing a significant amount of scattering in the ice. This sample is optimized for $\nu_\mu \rightarrow \nu_\mu$ disappearance channel by choosing $\nu_\mu$ CC events~\cite{IceCubeCollaboration:2023wtb}.
	
\begin{figure}
	\includegraphics[width=\linewidth]{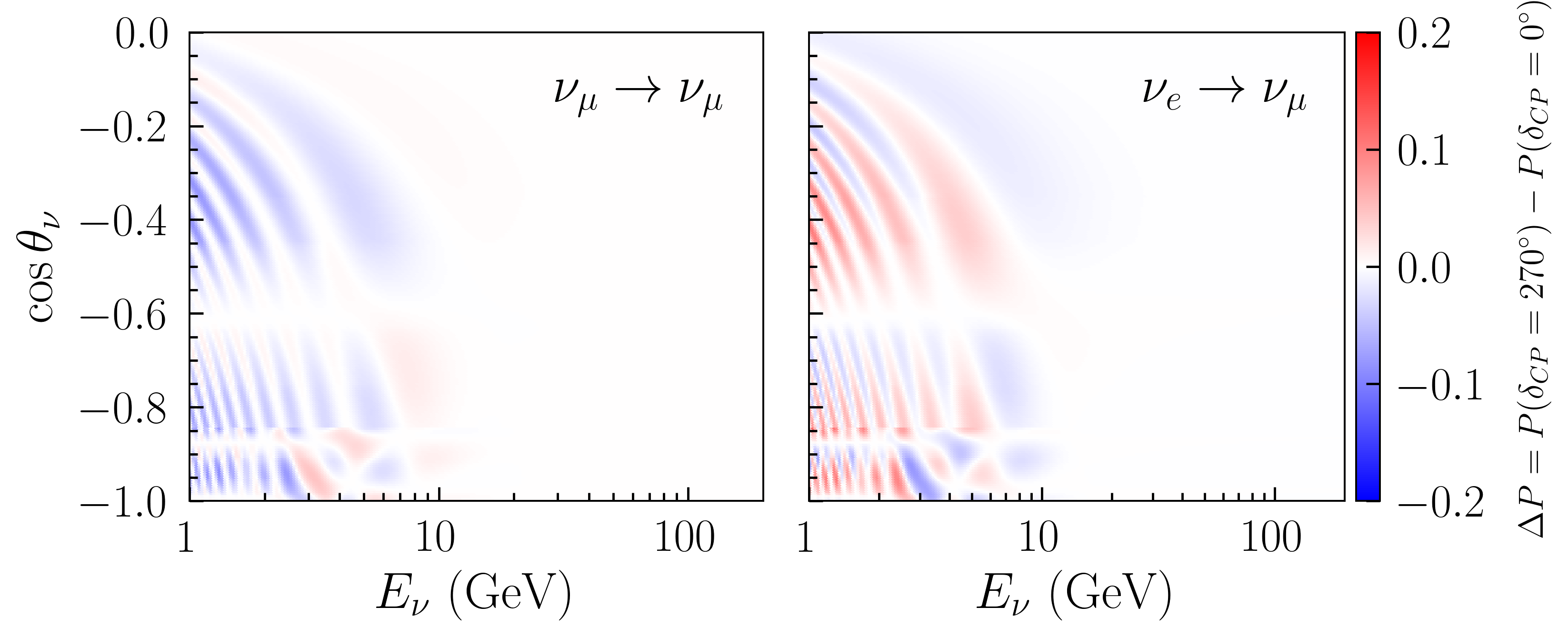}
	\caption{The differences in the three-flavor neutrino oscillation probabilities as a function of $E_\nu$ and $\cos\theta_\nu$ for the SI scenario with $\delta_{CP}=270^\circ$ and $\delta_{CP}=0^\circ$. The left panel shows the difference for the $\nu_{\mu}\rightarrow\nu_{\mu}$ channel, and the right panel shows the difference for the $\nu_e\rightarrow\nu_{\mu}$ channel.}
	\label{fig:cp_effect}
\end{figure}

The measurement of the NSI parameters $\varepsilon_{e\mu}$ and $\varepsilon_{e\tau}$ suffers from $\delta_{CP}$ degeneracy if data from appearance channels are used~\cite{Gago:2009ij,Liao:2016hsa,Ge:2016dlx,Agarwalla:2016fkh,Hyde:2018tqt}. On the other hand, the effect of $\delta_{CP}$ on $P(\nu_\mu \rightarrow \nu_\mu)$ is significantly suppressed by a factor of $(\Delta m^2_{21}/\Delta m^2_{31}) \times \sin\theta_{13} \sim 0.005$~\cite{Akhmedov:2004ny}. Figure~\ref{fig:cp_effect} shows the difference in the three-flavor neutrino oscillation probabilities for the standard interaction (SI) scenario between $\delta_{CP}=270^\circ$ and $\delta_{CP}=0^\circ$ for the $\nu_{\mu}\rightarrow\nu_{\mu}$ (left panel) and $\nu_{e}\rightarrow\nu_{\mu}$ (right panel) channels as a function of neutrino energy ($E_\nu$) and cosine of the zenith angle ($\cos\theta_{\nu}$). The small probability difference observed in Fig.~\ref{fig:cp_effect} indicate that the impact of $\delta_{CP}$ on $\nu_\mu\rightarrow\nu_\mu$ oscillation channel is negligible in the energy range of our interest. Therefore, the measurement of $\varepsilon_{e\mu}$ and $\varepsilon_{e\tau}$ using a high-purity $\nu_\mu$ CC sample comes with an advantage of being free from $\delta_{CP}$ degeneracy. These results would be complementary to those obtained from the long-baseline neutrino oscillation experiments, where the appearance channel depends significantly on the value of $\delta_{CP}$. The NSI parameters $\varepsilon_{\mu\tau}$ and $\varepsilon_{\tau\tau}-\varepsilon_{\mu\mu}$  have already been strongly constrained in our previous work~\cite{Krishnamoorthi:2025efw} using the same data sample. For comparison, those results are also included in this work for completeness.

This paper is organized as follows: In Sec.~\ref{sec:nsi_framework}, we briefly discuss the NSI framework and its effects on neutrino oscillation probabilities. In Sec.~\ref{sec:data_sample}, we describe the IceCube DeepCore detector and the atmospheric neutrino data sample used in this analysis. The analysis method, including the treatment of systematic uncertainties, is described in Sec.~\ref{sec:analysis}. The results of the analysis are presented in Sec.~\ref{sec:results}, where we discuss the constraints obtained for both flavor-conserving and flavor-violating NSI parameters. In Sec.~\ref{sec:Global_Comparison}, we compare the constraints obtained in this work with existing limits from other experiments. Finally, we summarize our findings and conclude in Sec.~\ref{sec:conclusion}. In Appendix~\ref{app:systematic_params}, we provide the best-fit values of various systematic parameters. We present the Data-MC comparison in Appendix~\ref{app:data_mc}.
	

\section{Effects of NSI on neutrino oscillation probabilities}\label{sec:nsi_framework}

The oscillations of atmospheric neutrinos passing through Earth depend not only on the neutrino energy and the baseline, but also on the additional matter potential arising from the CC coherent forward scattering of electron neutrinos with ambient electrons in the Earth. This matter potential depends on the electron number density ($N_e$) along the path. When we assume the presence of nonstandard interactions of neutrinos, the matter potential is further modified by the effective NSI parameters $\varepsilon_{\alpha\beta}$, which can affect the oscillation probabilities of neutrinos propagating through the Earth. The effective Hamiltonian in the flavor basis describing neutrino propagation in the presence of NSI can be written as
\begin{align}
	H_{\rm eff} &= \frac{1}{2E} \, U 
	\begin{pmatrix}
		0 & 0 & 0 \\
		0 & \Delta m^2_{21} & 0 \\
		0 & 0 & \Delta m^2_{31}
	\end{pmatrix} U^\dag \nonumber \\
	&\quad + V_{\rm CC}
	\begin{pmatrix}
		1 + \varepsilon_{ee} - \varepsilon_{\mu\mu} & \varepsilon_{e\mu} & \varepsilon_{e\tau} \\
		\varepsilon_{e\mu}^* & 0 & \varepsilon_{\mu\tau} \\
		\varepsilon_{e\tau}^* & \varepsilon_{\mu\tau}^* & \varepsilon_{\tau\tau} - \varepsilon_{\mu\mu}
	\end{pmatrix}\,,
	\label{eq:modified_H}
\end{align}
where $U$ is the PMNS mixing matrix~\cite{Maki:1962mu,Pontecorvo:1967fh}, $\Delta m^2_{21}$ and $\Delta m^2_{31}$ are the mass-squared differences, $E$ is the neutrino energy, and $V_{\rm CC} = \sqrt{2} G_F N_e$~\cite{Opher:1974drq, Langacker:1982ih} is the matter potential in the SI scenario. The diagonal NSI parameters $\varepsilon_{ee}-\varepsilon_{\mu\mu}$ and $\varepsilon_{\tau\tau}-\varepsilon_{\mu\mu}$ are real and flavor-conserving, whereas the off-diagonal parameters can be complex ($\varepsilon_{\alpha\beta}=|\varepsilon_{\alpha\beta}|e^{i\phi_{\alpha\beta}}$) and represent the flavor-violating NSI. For the case of antineutrinos, $U\rightarrow U^*$, $V_{\rm CC}\rightarrow -V_{\rm CC}$, and $\varepsilon_{\alpha\beta} \rightarrow \varepsilon_{\alpha\beta}^*$.


\begin{figure}[tp!]
	\includegraphics[width=\linewidth]{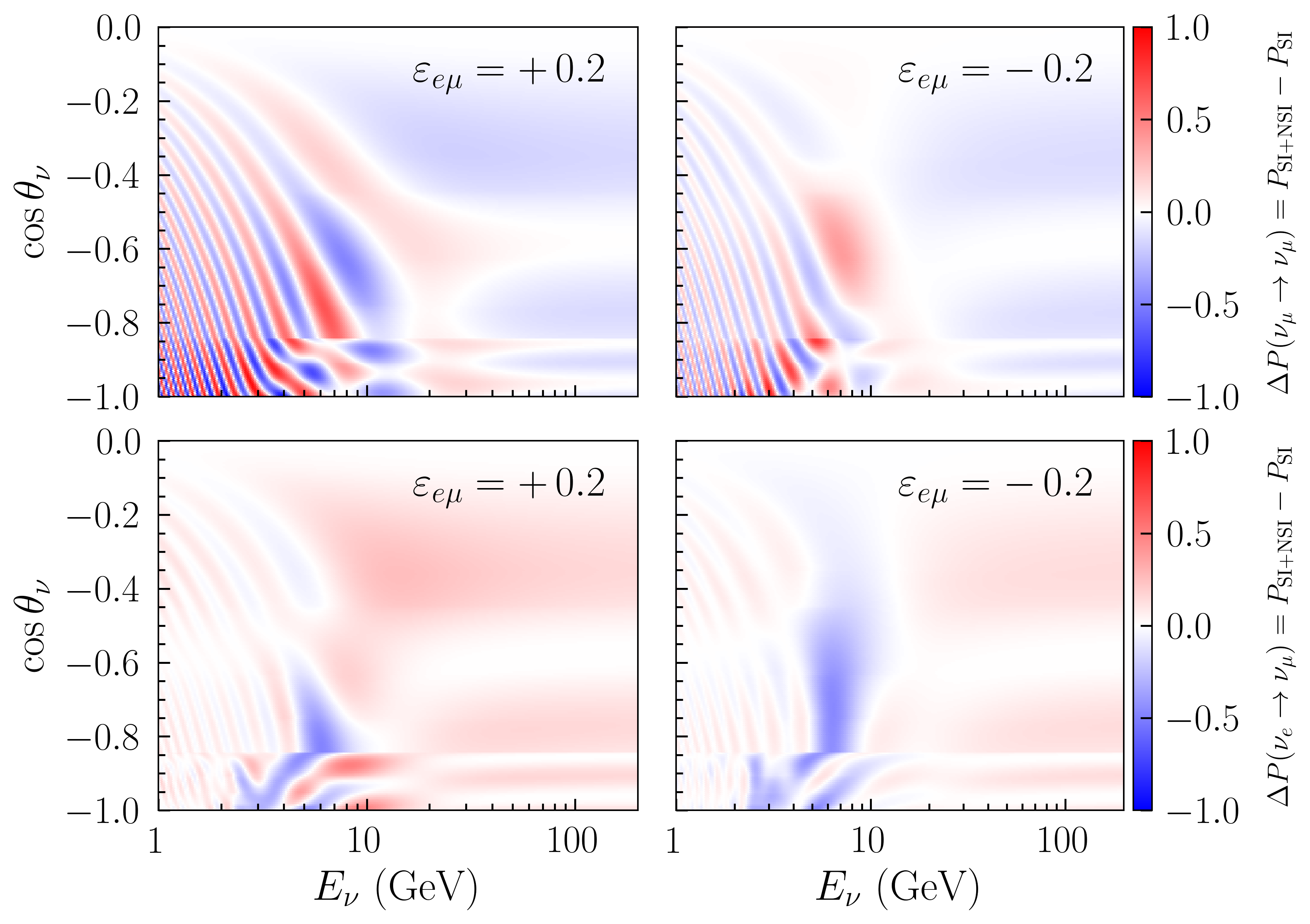}
	\caption{The differences of the three-flavor neutrino oscillation probabilities as a function of $E_\nu$ and $\cos\theta_{\nu}$ for the SI + NSI and SI scenarios, i.e., $\Delta P = P_{\rm SI+NSI} - P_{\rm SI}$. The left and right panels show the probability differences for the NSI parameter $\varepsilon_{e\mu}$  with the positive ($+\,0.2$) and negative ($-\,0.2$) values, respectively, assuming the other NSI parameters to be zero. The top panels show the difference for the $\nu_\mu \rightarrow \nu_\mu$ survival channel, and the bottom panels show that for the $\nu_e \rightarrow \nu_\mu$ appearance channel.}
	\label{fig:osc_dif_emu}
\end{figure}

\begin{figure}[tp!]
	\includegraphics[width=\linewidth]{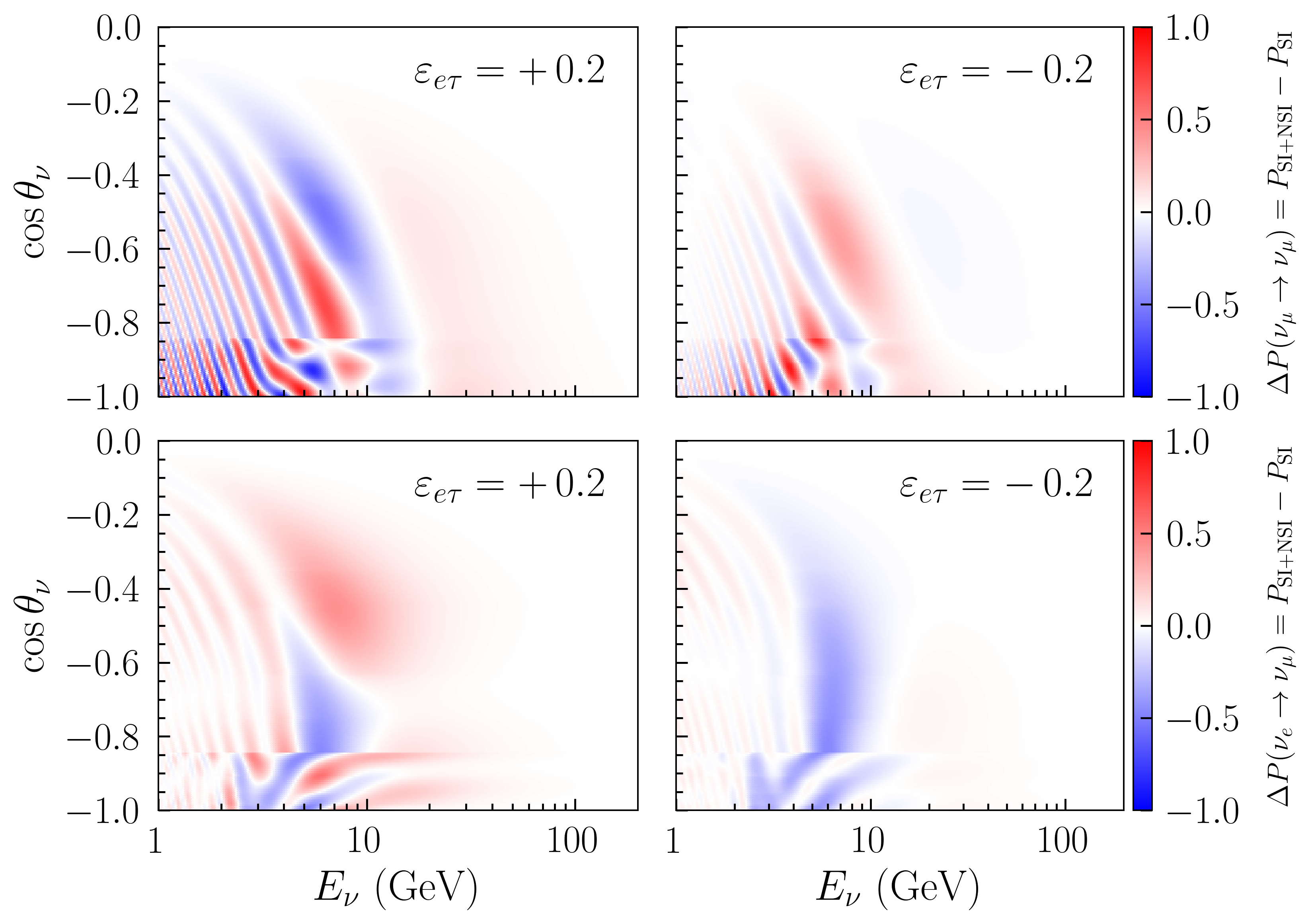}
	\caption{The differences of the three-flavor neutrino oscillation probabilities as a function of $E_\nu$ and $\cos\theta_{\nu}$ for the SI + NSI and SI scenarios, i.e., $\Delta P = P_{\rm SI+NSI} - P_{\rm SI}$. The left and right panels show the probability differences for the NSI parameter $\varepsilon_{e\tau}$ with the positive ($+\,0.2$) and negative ($-\,0.2$) values, respectively, assuming the other NSI parameters to be zero. The top panels show the difference for the $\nu_\mu \rightarrow \nu_\mu$ survival channel, and the bottom panels show that for the $\nu_e \rightarrow \nu_\mu$ appearance channel.}
	\label{fig:osc_dif_etau}
\end{figure}

\begin{figure}[htp!]
	\includegraphics[width=\linewidth]{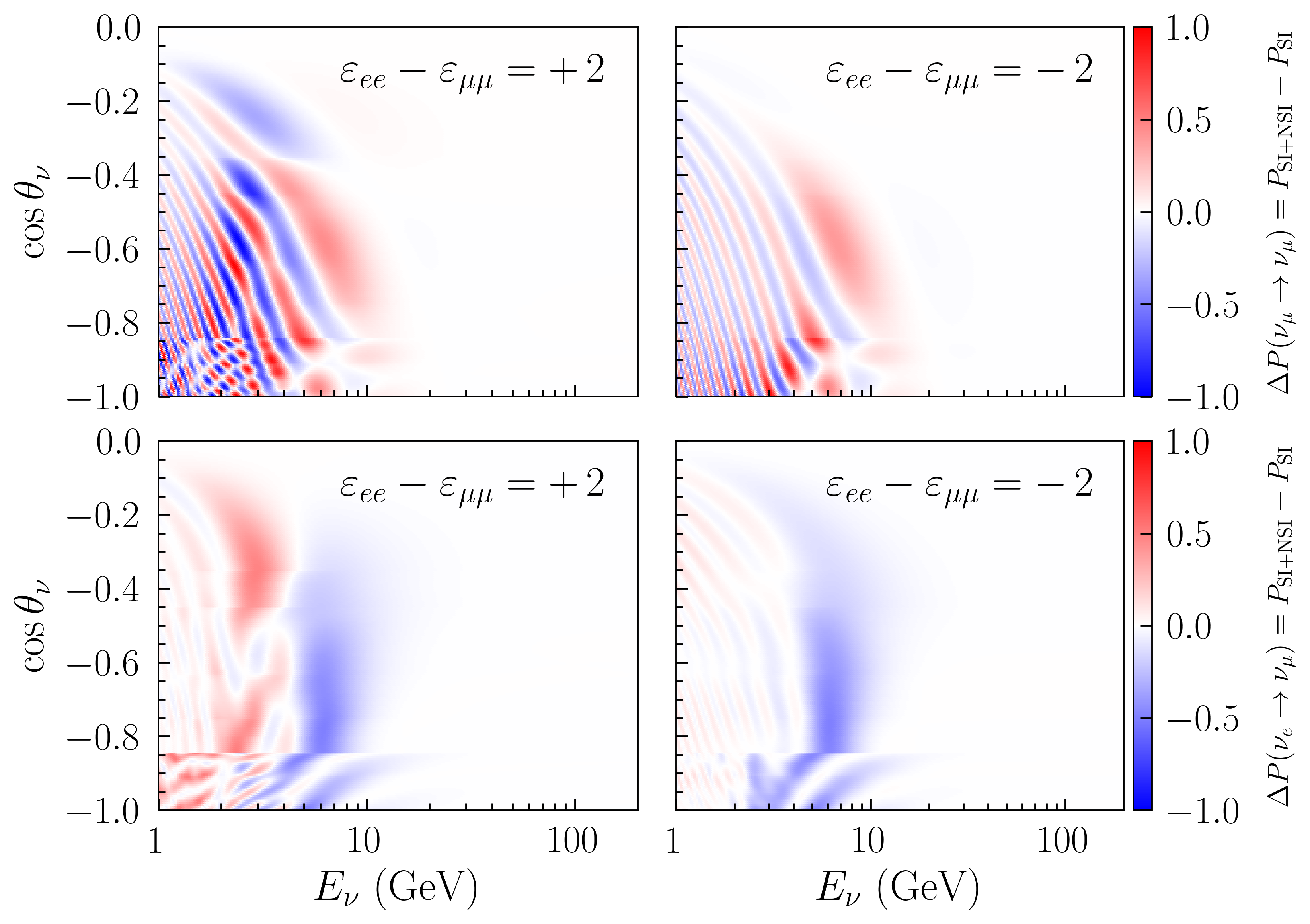}
	\caption{The differences of the three-flavor neutrino oscillation probabilities as a function of $E_\nu$ and $\cos\theta_{\nu}$ for the SI + NSI and SI scenarios, i.e., $\Delta P = P_{\rm SI+NSI} - P_{\rm SI}$. The left and right panels show the probability differences for the NSI parameter $\varepsilon_{ee}-\varepsilon_{\mu\mu}$ with the positive ($+\,2$) and negative ($-\,2$) values, respectively, assuming the other NSI parameters to be zero. The top panels show the difference for the $\nu_\mu \rightarrow \nu_\mu$ survival channel, and the bottom panels show that for the $\nu_e \rightarrow \nu_\mu$ appearance channel.}
	\label{fig:osc_dif_ee}
\end{figure}


Figures~\ref{fig:osc_dif_emu}, \ref{fig:osc_dif_etau}, and \ref{fig:osc_dif_ee} show the  difference between neutrino oscillation probabilities for the SI + NSI and SI scenarios. The differences of oscillation probabilities are shown as a function of the neutrino energy ($E_\nu$) and direction ($\cos\theta_{\nu}$) for the NSI parameters $\varepsilon_{e\mu}$, $\varepsilon_{e\tau}$, and $\varepsilon_{ee}-\varepsilon_{\mu\mu}$, respectively, considered one at a time. The top panels show the probability difference for the $\nu_\mu \rightarrow \nu_\mu$ survival channel, while the bottom panels show that for the $\nu_e \rightarrow \nu_\mu$ appearance channel. The left and right panels correspond to the positive and negative values of the NSI parameter under consideration, respectively, while assuming the other NSI parameters to be zero.

Figure~\ref{fig:osc_dif_emu}  (\ref{fig:osc_dif_etau}) shows that the NSI parameter $\varepsilon_{e\mu}$ ($\varepsilon_{e\tau}$) affects both $\nu_{\mu}\rightarrow\nu_{\mu}$ and $\nu_{e}\rightarrow{\nu_\mu}$ channels. However, the effects of $\varepsilon_{e\mu}$  ($\varepsilon_{e\tau}$) are more prominent in the $\nu_{e}\rightarrow{\nu_\mu}$ channel than in the $\nu_{\mu}\rightarrow\nu_{\mu}$. Although this is not immediately visible in Fig.~\ref{fig:osc_dif_emu} (Fig.~\ref{fig:osc_dif_etau}) as it presents absolute rather than relative probability differences. The parameters $\varepsilon_{e\mu}$ and $\varepsilon_{e\tau}$ show similar modifications to the oscillation probabilities, as they introduce new transitions $\nu_e\leftrightarrow \nu_\mu$ and $\nu_e\leftrightarrow\nu_\tau$, respectively. However, in the high-energy region ($E>20~\mathrm{GeV}$), the effect of $\varepsilon_{e\mu}$ is more than that of $\varepsilon_{e\tau}$. This is because $\varepsilon_{e\mu}$ directly affects the channel that involves $\nu_e$ and $\nu_\mu$ flavors. As far as the impact of $\varepsilon_{e\tau}$ at high energy is concerned, it appears in the $\nu_e\rightarrow\nu_\tau$ appearance channel.

To understand the effects of $\varepsilon_{e\mu}$ and $\varepsilon_{e\tau}$, we write the approximate expression for  $P(\nu_{e}\rightarrow{\nu_\mu})$ in matter in the presence of $|\varepsilon_{e\mu}|$ and $|\varepsilon_{e\tau}|$ with $\delta_{CP}=\phi_{e\mu}=\phi_{e\tau}=0$, following Eq.~(33) of Ref.~\cite{Kopp:2008}, where we retain terms having noticeable effects,
\begin{align}\label{eq:Peu_expression}
	&P(\nu_e \rightarrow \nu_\mu) \simeq
	4 \tilde{s}_{13}^{2} s_{23}^{2}
	\sin^{2} \frac{(\ldm - a_{\rm CC})L_\nu}{4E_\nu}
	\nonumber\\
	&- \alpha \tilde{s}_{13} s_{2\times12} s_{2\times23}
	\frac{\ldm}{a_{\rm CC}} \, \Delta_{\rm osc}
	\nonumber\\
	&- 4 |\varepsilon_{e\mu}| \tilde{s}_{13} s_{23} c_{23}^{2} \, \Delta_{\rm osc}
	\nonumber\\
	&+ 4 |\varepsilon_{e\tau}| \tilde{s}_{13} s_{23}^{2} c_{23} \, \Delta_{\rm osc}
	\nonumber\\
	&+ 8 |\varepsilon_{e\mu}| \tilde{s}_{13} s_{23}^{3}
	\frac{a_{\rm CC}}{\ldm - a_{\rm CC}}
	\sin^{2} \frac{(\ldm - a_{\rm CC})L_\nu}{4E_\nu}
	\nonumber\\
	&+ 8 |\varepsilon_{e\tau}| \tilde{s}_{13} s_{23}^{2} c_{23}
	\frac{a_{\rm CC}}{\ldm - a_{\rm CC}}
	\sin^{2} \frac{(\ldm - a_{\rm CC})L_\nu}{4E_\nu}
	\nonumber\\
	&+ \mathcal{O}(\alpha^3, \alpha^2, \alpha\varepsilon, \alpha s^2_{13}, s^3_{13}, \varepsilon^2)\,,
\end{align}
where
\begin{align}
	\alpha &\equiv \frac{\sdm}{\ldm}\,, \\
	\tilde{s}_{13} &\equiv \frac{\ldm}{\ldm - \,a_{\rm CC}} s_{13} + \mathcal{O}(s_{13}^2)\,, \\
	\Delta_{\rm osc} &\equiv
	\sin^2 \frac{a_{\rm CC} L_\nu}{4E_\nu}
	- \sin^2 \frac{\ldm L_\nu}{4E_\nu} \nonumber \\
	&+ \sin^2 \frac{(\ldm - a_{\rm CC})L_\nu}{4E_\nu}\,,
\end{align} 
$s_{ij} \equiv \sin\theta_{ij}$,
$c_{ij} \equiv \cos\theta_{ij}$,
$s_{2 \times ij} \equiv \sin 2\theta_{ij}$,
$c_{2 \times ij} \equiv \cos 2\theta_{ij}$,
and $a_{\rm CC} \equiv 2 V_{\rm CC} E_\nu$.

The first two terms in Eq.(~\ref{eq:Peu_expression}) represent the contribution from the standard interactions. The first term is the leading term corresponding to the atmospheric oscillations, and the second term is the subleading term representing the solar-atmospheric interference. The third and fifth terms contain the NSI contribution from $\varepsilon_{e\mu}$. While the third term modifies the same interference structure as the second term, the fifth term is additionally enhanced by the matter-driven resonance factor $a_{\rm CC}/(\ldm - a_{\rm CC})$. This factor can become large near the resonance condition $\Delta m^2_{31}\approx a_{\rm CC}$, making the fifth term the dominant contribution of $\varepsilon_{e\mu}$ in the regime driven by matter effects. The fourth and sixth terms account for the contribution of the parameter $\varepsilon_{e\tau}$, which produces similar effects on $P(\nu_e \rightarrow \nu_\mu)$ as observed for the case of $\varepsilon_{e\mu}$. These resonant features can also be observed in terms of the significant effects of $\varepsilon_{e\mu}$ and $\varepsilon_{e\tau}$ in the matter effect regions in Figs.~\ref{fig:osc_dif_emu} and \ref{fig:osc_dif_etau}, respectively.

Figure~\ref{fig:osc_dif_ee} shows the effects of $\varepsilon_{ee}-\varepsilon_{\mu\mu}$ parameter which is subleading below $10~\mathrm{GeV}$ in both  $P(\nu_{\mu}\rightarrow\nu_{\mu})$ and $P(\nu_{e}\rightarrow\nu_{\mu})$. As discussed in Ref.~\cite{Kikuchi:2008vq}, the parameter $\varepsilon_{ee}-\varepsilon_{\mu\mu}$ contributes to the oscillation probability only at the third order in the perturbative expansion. This parameter can also be interpreted as a modification in the standard matter potential. As a consequence, for positive values of $\varepsilon_{ee}-\varepsilon_{\mu\mu}$, the resonance energy shifts to lower values, which may fall below the detector threshold. For negative values of $\varepsilon_{ee}-\varepsilon_{\mu\mu}$, the resonance shifts to higher energies; however, when $\varepsilon_{ee}-\varepsilon_{\mu\mu}=-1$, the standard matter effects are canceled, and oscillations become similar to the vacuum oscillations. For larger negative values ($\varepsilon_{ee}-\varepsilon_{\mu\mu}<-1$), the sign of the effective matter potential reverses and the resonance condition occurs for antineutrinos even for normal mass ordering, whereas neutrinos show oscillations similar to vacuum oscillations. Because of this degeneracy with matter potential and the subleading contribution of this parameter, constraining $\varepsilon_{ee}-\varepsilon_{\mu\mu}$ is challenging. That is why we choose a relatively large benchmark value of $\varepsilon_{ee}-\varepsilon_{\mu\mu} = \pm\, 2$ for showing the effects of this parameter in Fig.~\ref{fig:osc_dif_ee}. The above-mentioned features can be seen in Fig.~\ref{fig:osc_dif_ee}, where the oscillation differences are relatively small and show a complex pattern. In the left panel, the Earth matter resonance is shifted toward lower energies. In contrast, in the right panel, the matter resonance is suppressed, effectively removing the matter effects. As a result, the observed difference is mainly due to the effective vacuumlike behavior of the NSI parameter $\varepsilon_{ee}-\varepsilon_{\mu\mu}$. In the next section, we discuss the data sample of IceCube DeepCore and the effects of the NSI parameters at the event level.

\section{IceCube DeepCore and data sample}\label{sec:data_sample}

The IceCube Neutrino Observatory~\cite{IceCube:2016zyt} is a cubic-kilometer-scale neutrino detector located inside the Antarctic ice at the South Pole. It is optimized to detect high-energy neutrinos coming from astrophysical sources. The central region of IceCube, called DeepCore, is more densely instrumented  and has digital optical modules (DOMs) with higher quantum efficiency~\cite{IceCube:2011ucd}. Along with 7 IceCube strings, DeepCore consists of 8 additional strings with closer spacing between DOMs compared to the main IceCube array. Because of this denser configuration and higher quantum efficiency of DOMs, DeepCore can detect neutrinos with energies of a few GeV. DeepCore is mainly optimized for studying atmospheric neutrinos that are produced during the interactions of primary cosmic rays with the nuclei of the Earth's atmosphere. In the present analysis, we search for NSI using publicly available atmospheric neutrino data collected by IceCube DeepCore during 2011 and 2018~\cite{DVN_B4RITM_2025}.

This data sample primarily includes those direct photon hits where scattering is not significant. Removing scattered photons improves the quality of events significantly, and hence, this sample is called the “golden event sample.” It has been processed with improved calibration and reconstruction techniques, leading to better energy and angular resolution compared to earlier samples. Details of the reconstruction techniques are described in Refs.~\cite{IceCube:2014flw,Garza2014Measurement,AndriiThesis,IceCube:2022kff}. The events in this data sample have reconstructed energies in the range of 6.3 GeV to 158.5 GeV and reconstructed zenith angles in the range $\cos\theta_z \in [-1, 0.1]$, corresponding to up-going events. The data is divided into topology-based categories: tracklike events, mixed events, and cascadelike events. This classification is based on the particle identification (PID) score obtained from a boosted decision tree (BDT)~\cite{Friedman:2001wbq}. The PID score corresponds to the probability of an event being $\nu_{\mu}$ CC-like. Events with PID in the range [0.75, 1.0] are classified as tracklike, those with PID in the range of [0.55, 0.75] are classified as mixed, and events with PID in the range of [0.0, 0.55] are classified as cascadelike. The cascadelike events are not included in the public sample by the IceCube Collaboration. The removal of cascadelike events and scattered photons increases the overall purity of $\nu_{\mu}$ CC interactions to $\sim 80\%$ in the data sample~\cite{IceCubeCollaboration:2023wtb}. In the next paragraph, we discuss the effect of NSI on event distributions. 

\begin{figure}[htp!]
	\includegraphics[width=\linewidth]{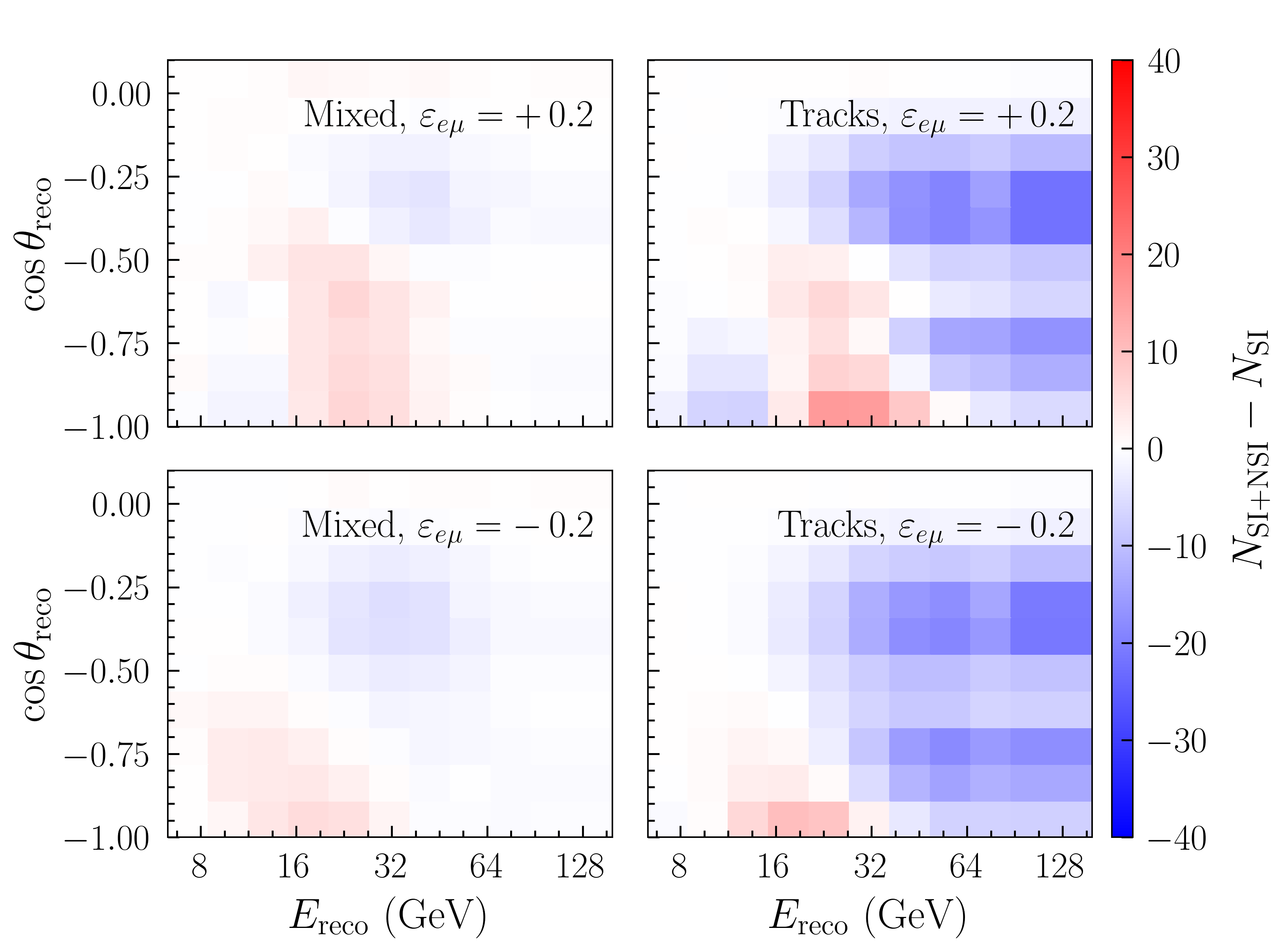}
	\caption{Difference of expected event distributions between the presence of NSI ($N_{\rm SI+NSI}$) and the standard interaction ($N_{\rm SI}$) case as a function of reconstructed energy and cosine of zenith angle. The top and bottom panels show event differences for $\varepsilon_{e\mu} = +\,0.2$ and $-\,0.2$, respectively. The left and right panels represent mixed and tracklike events, respectively.}
	\label{fig:event_dif_emu}
\end{figure}

\begin{figure}[htp!]
	\includegraphics[width=\linewidth]{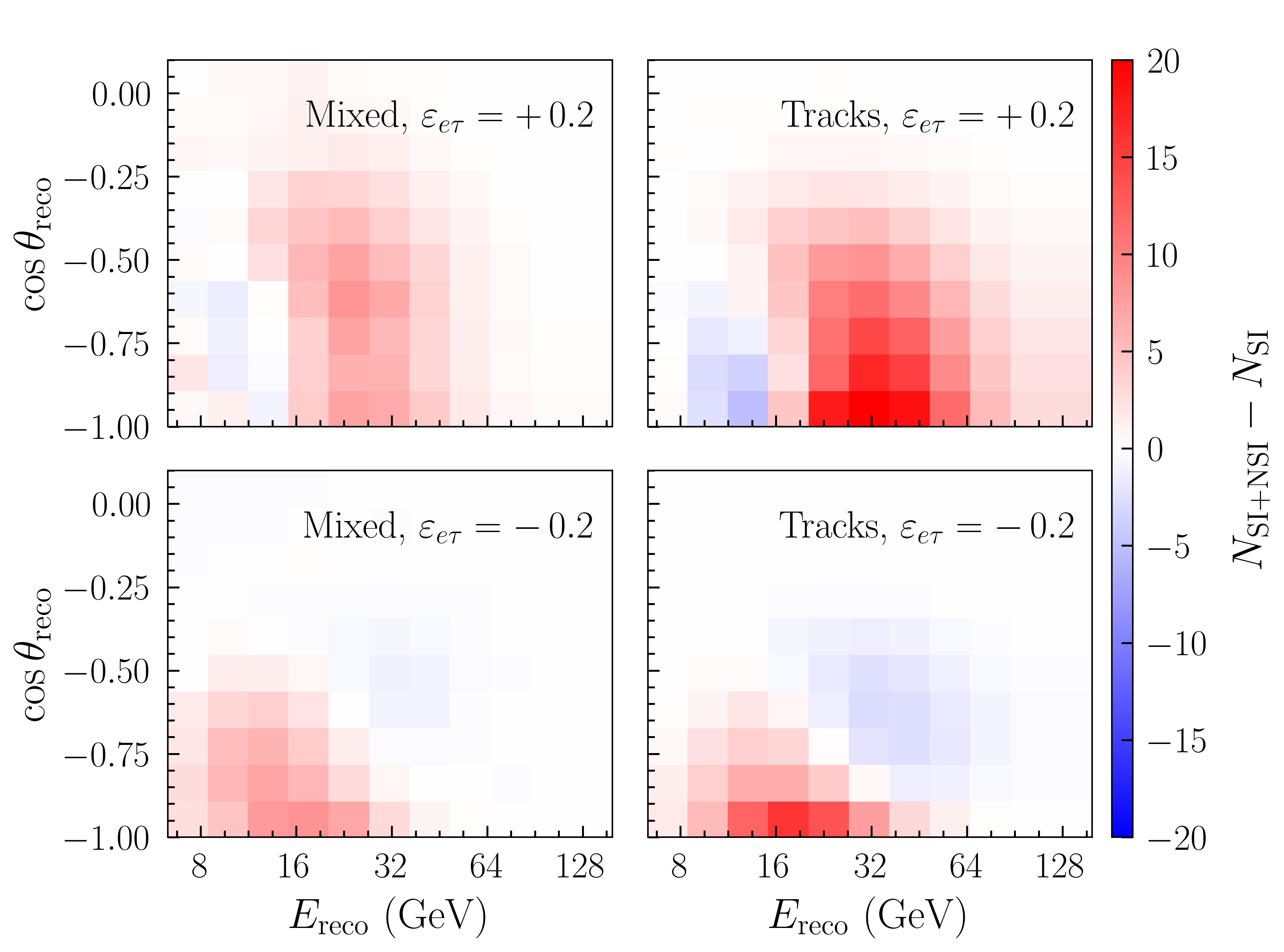}
	\caption{Difference of expected event distributions between the presence of NSI ($N_{\rm SI+NSI}$) and the standard interaction ($N_{\rm SI}$) case as a function of reconstructed energy and cosine of zenith angle. The top and bottom panels show event differences for $\varepsilon_{e\tau} = +\,0.2$ and $-\,0.2$, respectively. The left and right panels represent mixed and tracklike events, respectively.}
	\label{fig:event_dif_etau}
\end{figure}

\begin{figure}[htp!]
	\includegraphics[width=\linewidth]{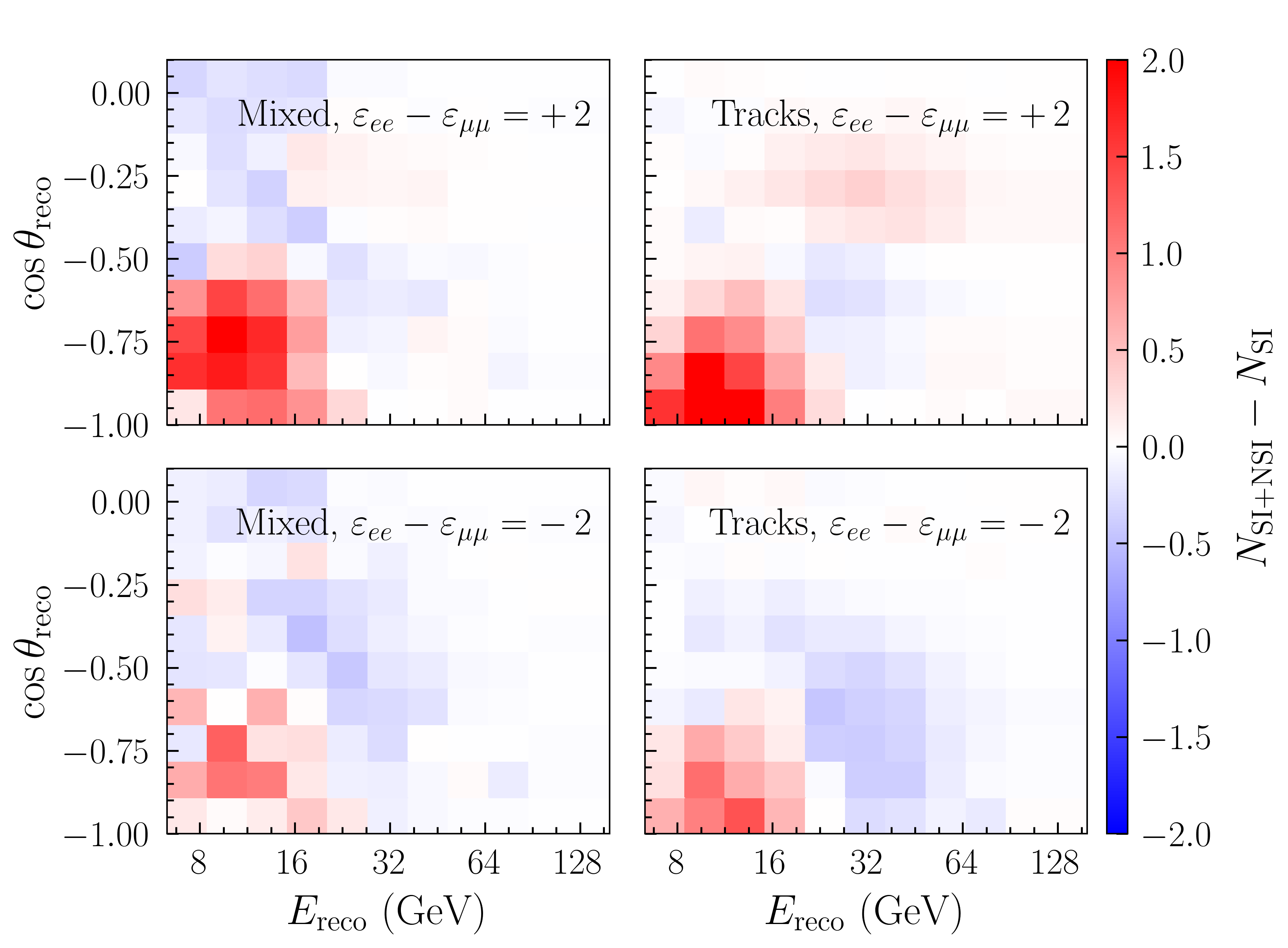}
	\caption{Difference of expected event distributions between the presence of NSI ($N_{\rm SI+NSI}$) and the standard interaction ($N_{\rm SI}$) case as a function of reconstructed energy and cosine of zenith angle. The top and bottom panels show event differences for $\varepsilon_{ee}-\varepsilon_{\mu\mu} = +\,2$ and $-\,2$, respectively. The left and right panels represent mixed and tracklike events, respectively.}
	\label{fig:event_dif_ee}
\end{figure}

Figures~\ref{fig:event_dif_emu}, \ref{fig:event_dif_etau} and~\ref{fig:event_dif_ee}, represent the difference of expected event distributions as a function of reconstructed energy ($E_{\rm reco}$) and reconstructed cosine of zenith angle ($\cos\theta_{\rm reco}$) for the NSI parameters $\varepsilon_{e\mu}$, $\varepsilon_{e\tau}$, and $\varepsilon_{ee}-\varepsilon_{\mu\mu}$,  respectively, considered one at a time. The event differences are calculated between the number of events in the presence of NSI ($N_{\rm SI+NSI}$) and the number of events in the standard interaction ($N_{\rm SI}$) case. The left and right panels represent the mixed and tracklike events. The energy range has 10 logarithmically spaced bins, and the cosine of zenith angle range has 10 linearly spaced bins. The top and bottom panels show event differences for the positive and negative values of the NSI parameter under consideration, respectively, while taking the other NSI parameters to be zero. These event difference distributions highlight the signal regions and illustrate the characteristic patterns of deviations in the reconstructed energy and cosine of zenith angle due to the presence of NSI.
	
Figure~\ref{fig:event_dif_emu} shows the event difference corresponding to $\varepsilon_{e\mu}$, where the signal extends to the higher energy region, which is similar to the features observed in the oscillation probability difference in Fig.~\ref{fig:osc_dif_emu}. Here, both positive and negative values of $\varepsilon_{e\mu}$ show similar event difference patterns. Figure~\ref{fig:event_dif_etau} shows the signal in the mid-energy region for the parameter $\varepsilon_{e\tau}$ as seen in the oscillation probability difference in Fig.~\ref{fig:osc_dif_etau}. The amplitude of the event difference for $\varepsilon_{e\tau}$ is relatively lower compared to that for $\varepsilon_{e\mu}$. Figure~\ref{fig:event_dif_ee} corresponding to $\varepsilon_{ee}-\varepsilon_{\mu\mu}$ shows the signal in the low energy and high baseline regions, which is consistent with the oscillation probability difference shown in Fig.~\ref{fig:osc_dif_ee}. Note that the event differences are lower than those in Figs.~\ref{fig:event_dif_emu} and~\ref{fig:event_dif_etau}.

\section{Analysis methodology}\label{sec:analysis}

In this work, we perform a binned $\chi^2$ analysis, where we compare the observed data with the expected event distributions under specific NSI hypothesis considering one parameter at a time. The binned $\chi^2$ function~\cite{IceCube:2017lak} is defined as
\begin{equation}
	\chi^2_{\rm mod} = \sum_{i \in {\rm bins}} \frac{(N_i^{\rm exp} - N_i^{\rm obs})^2}{N_i^{\rm exp} + (\sigma_i^{\rm sim})^2} + \sum_{j \in {\rm syst}} \frac{(s_j - \hat{s}_j)^2}{\sigma^2_{s_j}}\,,
	\label{eqn:mod_chi2}
\end{equation}
where $N_i^{\rm exp}$ and $N_i^{\rm obs}$ are the expected and observed number of events in the $i$th bin, respectively. Here, $\sigma_i^{\rm sim}$ represents the statistical uncertainty associated with the expected event counts due to the limited Monte Carlo statistics. The second summation accounts for the systematic uncertainties, where $s_j$ is the value of the $j$th systematic parameter, $\hat{s}_j$ is its nominal value, and $\sigma_{s_j}$ is the corresponding uncertainty.

The analysis has two main objectives: to fit the observed data to determine the best-fit values of the NSI parameters and to derive constraints on them. While fitting the observed data, the best-fit values are obtained by minimizing the $\chi^2$ function with respect to one NSI parameter at a time along with all nuisance parameters included in the analysis. This minimized $\chi^2$ is denoted as $\chi^2(\rm all \,\,free)$. To derive constraints on a given NSI parameter, we fit the observed data by performing a $\chi^2$ scan where that NSI parameter is fixed at different values in theory and $\chi^2$ is minimized with respect to the nuisance parameters. This minimized $\chi^2$ at a fixed NSI parameter in theory is denoted as $\chi^2(\text{NSI fixed})$. The test statistic for the analysis is defined as 
\begin{equation}
\Delta \chi^2=\chi^2(\text{NSI fixed})-\chi^2(\text{all\,\,free})\,.
\end{equation}
The constraints on the NSI parameter are then obtained from the $\Delta \chi^2$ values corresponding to the desired confidence level for the appropriate number of degrees of freedom.

The fitting in the analysis involves a detailed treatment of systematic uncertainties following Ref.~\cite{IceCubeCollaboration:2023wtb}. We incorporate the systematic uncertainties related to the detector, atmospheric flux, neutrino-nucleon interaction cross section, neutrino oscillation parameters, and background, such as atmospheric muons. There are a total of 20 nuisance parameters considered in the analysis, which are freely varied during the $\chi^2$ minimization.

\begin{itemize}
	
	\item \textit{Detector:} The detector-related uncertainties are parameterized by five nuisance parameters. This includes a global efficiency parameter for the DOMs and two parameters each for the optical properties corresponding to scattering and absorption of photons in ice. Two additional nuisance parameters are included to account for the effects of refrozen ice in the string holes on the angular acceptance of the DOMs.
	
	\item \textit{Atmospheric flux:} Seven nuisance parameters are used to describe the atmospheric neutrino flux model, which includes the uncertainties in the spectral index and the production of charged mesons in the atmosphere~\cite{Barr:2006it}.
	
	\item \textit{Cross section:} The neutrino-nucleon interaction cross-section uncertainties are parameterized by four nuisance parameters, which include the axial mass parameter for charged-current quasi-elastic interactions, the axial mass parameter for charged-current resonant interactions, a single scaling parameter for deep inelastic scattering, and the ratio of neutral-current to charged-current interaction cross sections.
	
	\item \textit{Oscillation parameters:} The atmospheric neutrino oscillation parameters $\theta_{23}$ and $\Delta m_{31}^2$ are included as nuisance parameters. Other oscillation parameters, except $\delta_{CP}$, are kept fixed at the values reported by NuFit v5.2~\cite{Esteban:2020cvm}. Throughout this work, we assume normal mass ordering. We have checked that the impact of taking $\delta_{CP}$ as a free nuisance parameter is negligible on this analysis. We have also checked that the results do not depend on the choice of the true value of $\delta_{CP}$. Therefore, $\delta_{CP}$ is fixed at $0^\circ$ during the fit. These checks ensure that the NSI constraints presented in this work are free from $\delta_{CP}$ degeneracy.
	
	\item \textit{Normalization:} Two separate normalization parameters are included for neutrinos and atmospheric muons.
\end{itemize}

The list of all the nuisance parameters included in the analysis, along with their nominal values and uncertainties, is summarized in Appendix~\ref{app:systematic_params}. The effects of these nuisance parameters are incorporated in the analysis by reweighting the expected event distributions according to the variations in these parameters. The reweighting of expected events and the minimization of $\chi^2$ are performed using the open source analysis framework \texttt{PISA}~\cite{IceCube:2018ikn}, which is provided by the IceCube Collaboration.


\section{Results}\label{sec:results}

In this section, we present the results of the analysis where we use the 8-year golden event sample of IceCube DeepCore to search for the flavor-violating NSI parameters $\varepsilon_{e\mu}$ and $\varepsilon_{e\tau}$, and the flavor-conserving (nonuniversal)  NSI parameter $\varepsilon_{ee}-\varepsilon_{\mu\mu}$. These NSI parameters are probed one at a time, where the other NSI parameters are kept fixed at zero. The summary of the best-fit values of the NSI parameters is reported in TabLE~\ref{tab:best-fit-table}, along with the corresponding $\Delta\chi^2_{\rm SI-NSI}$ values relative to the standard interaction case. For the completeness of the comparison, the best-fit values and the $\Delta\chi^2$ values relative to the standard interaction case for the parameter $\varepsilon_{\mu\tau}$ and $\varepsilon_{\tau\tau}-\varepsilon_{\mu\mu}$ are also included from our previous work~\cite{Krishnamoorthi:2025efw}. In all the cases, we do not observe any significant deviation from the standard interaction case. Since the data sample is found to be consistent with no NSI hypothesis, we place constraints on the NSI parameters, which are compared with the expected sensitivities, as well as with the previous IceCube DeepCore results~\cite{IceCubeCollaboration:2021euf} from a 3-year data sample. The bounds obtained in this analysis are consistent with the previous IceCube DeepCore results and provide complementary constraints to those from the other experiments.

\begin{table}[htp!]
	\centering
	\renewcommand{\arraystretch}{1.2}
	\begin{tabular}{l@{\hskip 15pt}c@{\hskip 15pt}c}
		\hline
		\hline
		Parameters                                     & Best-fit values                                           & $\Delta \chi^2_\text{SI|NSI}$ \\
		\hline
		$\varepsilon_{e\mu}$                          & $|\varepsilon_{e\mu}|=0.034$, $\phi_{e\mu}= 346.2^{\circ}$  & 0.18                          \\
		$\varepsilon_{e\tau}$                         & $|\varepsilon_{e\tau}|=0.11$, $\phi_{e\tau}=8.8^{\circ}$ & 1.31                          \\
		$\varepsilon_{ee}-\varepsilon_{\mu\mu}$       & $-0.59$                                                   & 0.25                          \\
		$\varepsilon_{\mu\tau}$~\cite{Krishnamoorthi:2025efw}                       & $|\varepsilon_{\mu\tau}|=0.0014$, $\phi_{\mu\tau}=178.8^{\circ}$  & 0.06                          \\
		$\varepsilon_{\tau\tau}-\varepsilon_{\mu\mu}$~\cite{Krishnamoorthi:2025efw} & $-0.0009$                                                 & 0.0017                        \\	
		\hline
		\hline
	\end{tabular}
	\caption{Best-fit values of the NSI parameters obtained after fitting the 8-year golden event sample of IceCube DeepCore, along with the corresponding $\Delta\chi^2$ relative to the standard interaction case. For comparison, the best-fit values and $\Delta\chi^2$ for the parameters $\varepsilon_{\tau\tau}-\varepsilon_{\mu\mu}$ and $\varepsilon_{\mu\tau}$ from our previous work~\cite{Krishnamoorthi:2025efw} are also included.}
	\label{tab:best-fit-table}
\end{table}

The following subsections discuss the bounds obtained for the NSI parameters $\varepsilon_{e\mu}$, $\varepsilon_{e\tau}$, and $\varepsilon_{ee}-\varepsilon_{\mu\mu}$. The bounds are reported at 90\% confidence level (CL) for the two degrees of freedom (2 DOF) for $\varepsilon_{e\mu}$ and $\varepsilon_{e\tau}$, and at 90\% CL for one degree of freedom (1 DOF) for $\varepsilon_{ee}-\varepsilon_{\mu\mu}$. In addition to these, the flavor-violating parameters $\varepsilon_{e\mu}$ and $\varepsilon_{e\tau}$ are also taken to be real by fixing their phases to 0 and $\pi$ for positive and negative values, respectively, and the bounds are derived at 90\% CL assuming 1 degree of freedom.

\subsection{Constraints on $\varepsilon_{e\mu}$}

\begin{figure*}[htp!]
	\centering
	\includegraphics[width=\textwidth]{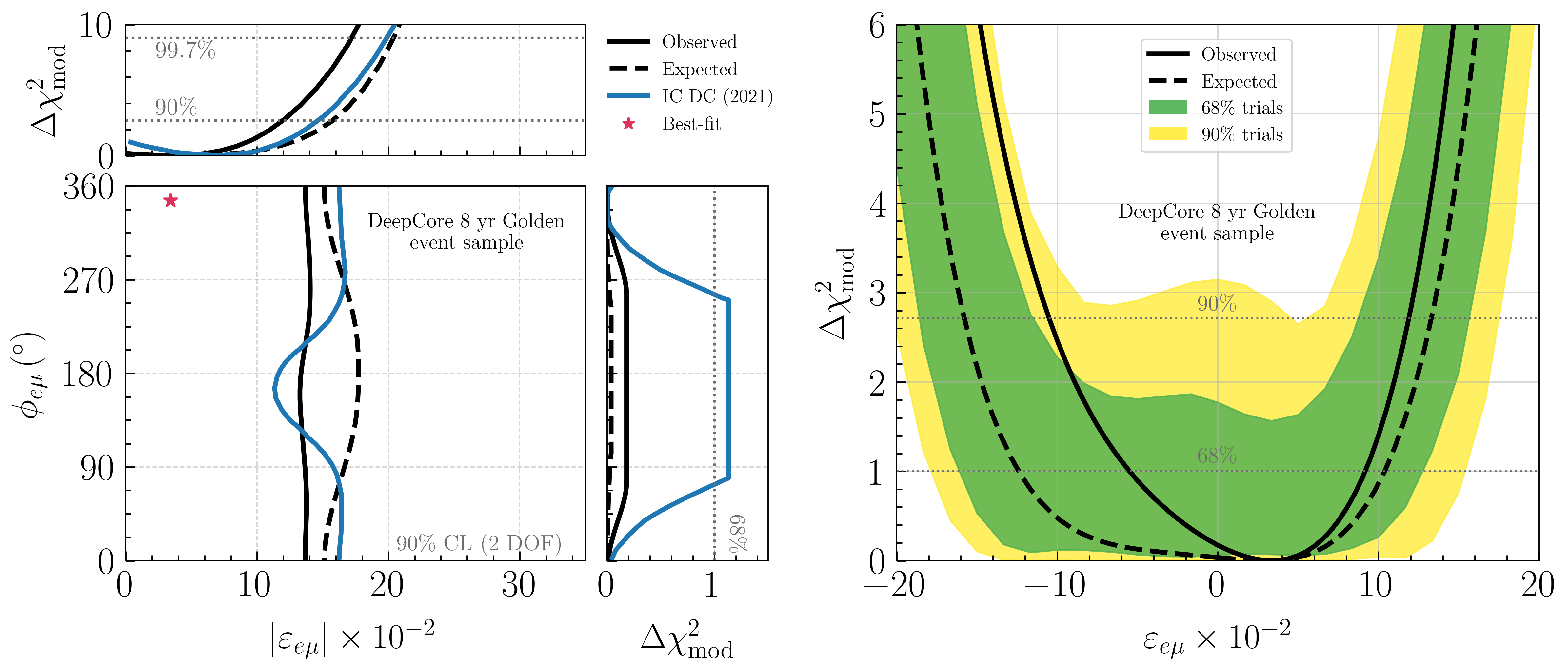}
	\caption{Left: the observed (solid curve) and expected (dashed curve) contours at 90\% CL (2 degrees of freedom) for the magnitude $|\varepsilon_{e\mu}|$ and phase $\phi_{e\mu}$. The red star marker denotes the best-fit values of the magnitude and phase. The top (side) subpanel represents the 1D projection of the 2D contour for the magnitude (phase), where the minimization is performed over the phase (magnitude) and nuisance parameters. The blue curves represent the $\Delta \chi^2$ from the IceCube DeepCore results using the 3-year data sample~\cite{IceCubeCollaboration:2021euf}. The dotted lines in the subpanels show the 68.3\%, 90\%, and 99.7\% confidence levels for 1 degree of freedom. Right: the observed (solid curve) and expected (dashed curve) $\Delta \chi^2$ profiles for the NSI parameter $\varepsilon_{e\mu}$, assuming the parameter to be real by fixing its complex phase to either $0$ (positive) or $\pi$ (negative). The color bands represent the range of expected $\Delta \chi^2$ when the fit is performed over statistically fluctuated pseudoexperiments. The horizontal dotted lines show the 68.3\% and 90\% confidence levels for 1 degree of freedom.}
	\label{fig:emu_result}
\end{figure*}

The left panel of Fig.~\ref{fig:emu_result} shows the allowed regions at 90\% CL (2 DOF) for the magnitude and phase of the NSI parameter $\varepsilon_{e\mu}$. The solid contour denotes the observed constraints. The dashed contour represents the expected sensitivity using the simulated data as truth that uses the best-fit values obtained from data fitting. For comparison, the IceCube DeepCore result~\cite{IceCubeCollaboration:2021euf} based on the 3-year data sample is also included as the blue solid curve. The top (side) subpanel illustrates the one-dimensional projection of the magnitude (phase) of $\varepsilon_{e\mu}$, which is obtained by profiling over the phase (magnitude) and nuisance parameters. The dotted gray lines in these subpanels correspond to the 68.3\%, 90\%, and 99.7\% confidence levels for 1 DOF.

The best-fit values for the magnitude and phase of $\varepsilon_{e\mu}$ are $0.034$ and $346.2^\circ$, respectively, with a p-value of 0.26. We observe no significant deviation from the standard interaction case, with $\Delta\chi^2_{\rm SI-NSI}  = 0.18$. We obtain an upper bound of $|\varepsilon_{e\mu}| \leq 0.12$ at 90\% confidence level. No constraint can be obtained on the phase, because minimization over the magnitude always prefers its values close to zero, giving a negligible $\Delta\chi^2$. Even though the present analysis uses relatively lower statistics, the constraints obtained here are comparable and consistent with the previous IceCube DeepCore results. This is because the present sample includes clean events with high $\nu_\mu$ CC purity, which also provides constraints on these parameters that are free from degeneracy with $\delta_{CP}$. 

The right panel of Fig.~\ref{fig:emu_result} shows the constraints on $\varepsilon_{e\mu}$ assuming real values, that are obtained by fixing the phase to $0$ for positive values and $\pi$ for negative values. The observed and expected $\Delta\chi^2$ profiles, shown as solid and dashed black curves, respectively, are in good agreement. The expected profile is obtained using the simulated data as truth, where the best-fit values from the fit to observed data are used. The horizontal gray dotted lines indicate the 68.3\% and 90\% confidence levels for 1 degree of freedom. The observed allowed range for $\varepsilon_{e\mu}$ at 90\% CL is $[-0.11\,,0.12]$, which is consistent with the expected sensitivity. The green (yellow) band represent the 68.3\% (90\%) range of expected $\Delta\chi^2$ that are obtained by fitting the statistically fluctuated pseudoexperiments. The observed $\Delta\chi^2$ profile lies within these bands, which indicates consistency with statistical expectations.

\subsection{Constraints on $\varepsilon_{e\tau}$}

\begin{figure*}[htp!]
	\centering
	\includegraphics[width=\textwidth]{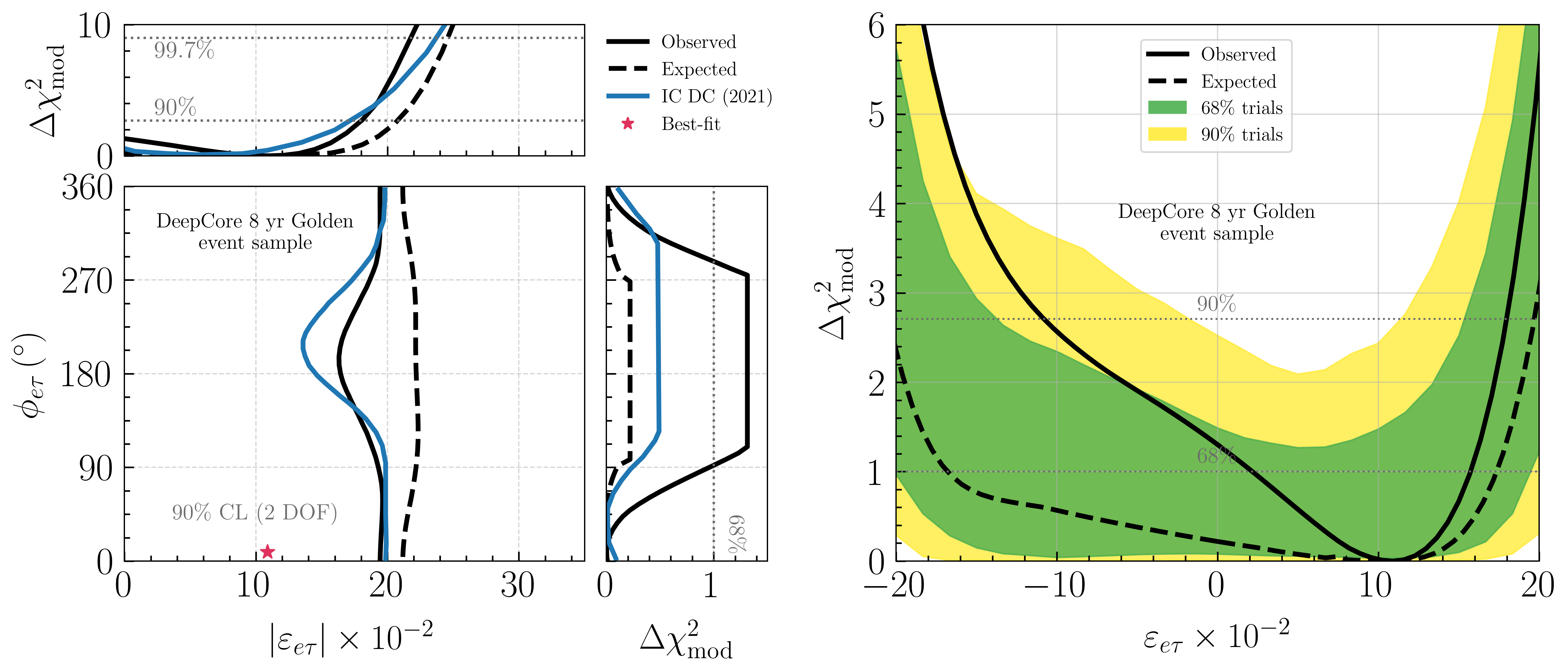}
	\caption{Left: the observed (solid curve) and expected (dashed curve) contours at 90\% CL (2 degrees of freedom) for the magnitude $|\varepsilon_{e\tau}|$ and phase $\phi_{e\tau}$. The red star marker denotes the best-fit values of the magnitude and phase. The top (side) subpanel represents the 1D projection of the 2D contour for the magnitude (phase), where the minimization is performed over the phase (magnitude) and nuisance parameters. The blue curves represent the $\Delta \chi^2$ from the IceCube DeepCore results using the 3-year data sample~\cite{IceCubeCollaboration:2021euf}. The dotted lines in the subpanels shows the 68.3\%, 90\%, and 99.7\% confidence levels for 1 degree of freedom. Right: the observed (solid curve) and expected (dashed curve) $\Delta \chi^2$ profiles for the NSI parameter $\varepsilon_{e\tau}$, assuming the parameter to be real by fixing its complex phase to either $0$ (positive) or $\pi$ (negative). The color bands represent the range of expected $\Delta \chi^2$ when the fit is performed over statistically fluctuated pseudoexperiments. The horizontal dotted lines show the 68.3\% and 90\% confidence levels for 1 degree of freedom.}
	\label{fig:etau_results}
\end{figure*}

The left panel of Fig.~\ref{fig:etau_results} presents the allowed regions at 90\% CL (2 DOF) for the magnitude and phase of the NSI parameter $\varepsilon_{e\tau}$. The solid contour denotes the observed limits. The dashed contour represents the expected sensitivity, where the simulated data is used as truth using the best-fit values obtained from fitting the observed data. For comparison, the IceCube DeepCore result~\cite{IceCubeCollaboration:2021euf} using the 3-year data sample is also included as the blue solid curve. The top (side) subpanel illustrates the one-dimensional projection for the magnitude (phase), where the minimization is performed over the phase (magnitude) and nuisance parameters. The dotted lines in these subpanels correspond to the 68.3\%, 90\%, and 99.7\% confidence levels for 1 DOF.

The best-fit values of the magnitude and phase of $\varepsilon_{e\tau}$ are $0.11$ and $8.8^\circ$, respectively, with a p-value of 0.27. The best fit is compatible with the standard interaction hypothesis with $\Delta\chi^2_{\rm SI-NSI} = 1.31$. We obtain an upper bound of $|\varepsilon_{e\tau}| \leq 0.18$ at 90\% confidence level. We are not able to constrain the phase at 90\% CL, because minimization over the magnitude always prefers its values close to zero, giving very small $\Delta\chi^2$.

Although this sample has twice the livetime of the earlier DeepCore analysis~\cite{IceCubeCollaboration:2021euf}, the sensitivity improvement is small because the event selection is optimized for $\nu_\mu$ CC events, which reduces the statistics. Since $\varepsilon_{e\tau}$ is mainly constrained through appearance channels rather than $\nu_\mu \rightarrow \nu_\mu$, the improvement in the bound is limited. However, the advantage of using $\nu_\mu \rightarrow \nu_\mu$ channel is that the constraints obtained on $\varepsilon_{e\tau}$  are free from $\delta_{CP}$ degeneracy.

The right panel of Fig.~\ref{fig:etau_results} shows the constraints on $\varepsilon_{e\tau}$ assuming real values, that are obtained by fixing the phase to $0$ (positive $\varepsilon_{e\tau}$) and $\pi$ (negative $\varepsilon_{e\tau}$). The observed and expected $\Delta\chi^2$ profiles, shown as solid and dashed curves, respectively, are in good agreement. Here, the expected  $\Delta\chi^2$ profile is obtained using simulated data as truth, where the best-fit values from the fit to the observed data are used. The horizontal dotted lines indicate the 68.3\% and 90\% confidence levels for 1 degree of freedom. The observed 90\% CL allowed range for $\varepsilon_{e\tau}$ is $[-0.11\,,0.18]$, which is consistent with expectations. The green (yellow) band represents the 68.3\% (90\%) range for expected $\Delta\chi^2$, where fit is performed over statistically fluctuated pseudoexperiments. The observed $\Delta\chi^2$ profile lies within these colored bands, which indicates consistency with statistical expectations.

\subsection{Constraints on $\varepsilon_{ee}-\varepsilon_{\mu\mu}$}

\begin{figure}[htp!]
	\includegraphics[width=\linewidth]{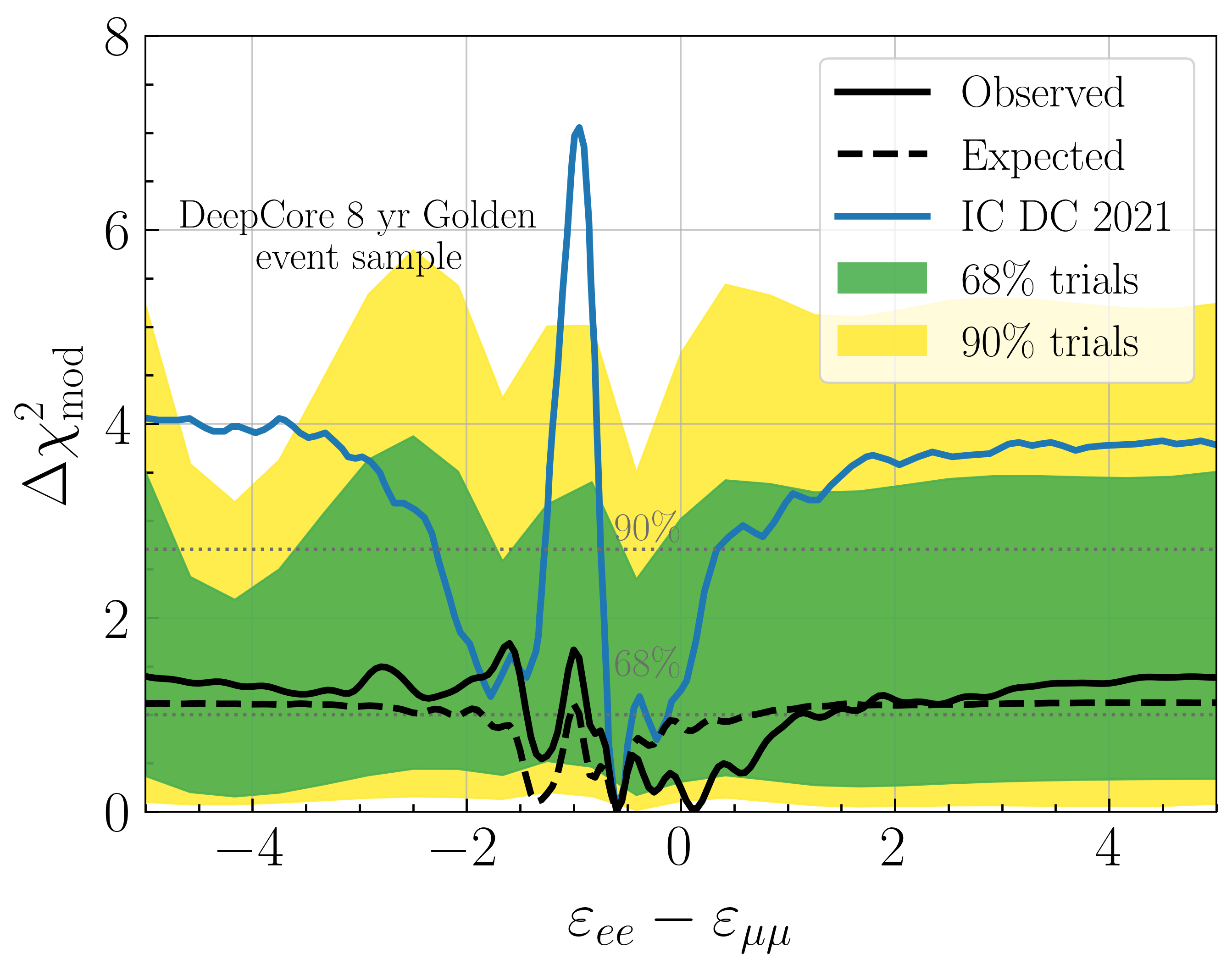}
	\caption{The observed (solid curve) and expected (dashed curve) $\Delta \chi^2$ profiles for the NSI parameter $\varepsilon_{ee} - \varepsilon_{\mu\mu}$. The blue curve represents the $\Delta \chi^2$ profile from the IceCube DeepCore result using the 3-year data sample~\cite{IceCubeCollaboration:2021euf}. The color bands correspond to the range of expected $\Delta \chi^2$ when the fit is performed over statistically fluctuated pseudoexperiments. The horizontal dotted lines show the 68.3\% and 90\% confidence levels for 1 degree of freedom.}
	\label{fig:diag_scan}
\end{figure}

Figure~\ref{fig:diag_scan} shows the $\Delta\chi^2$ profile for the nonuniversal (diagonal) NSI parameter $\varepsilon_{ee}-\varepsilon_{\mu\mu}$ obtained in this analysis. For comparison, we have also shown the corresponding result from the previous IceCube DeepCore measurement using 3 years of data~\cite{IceCubeCollaboration:2021euf}. The two $\Delta\chi^2$ profiles are consistent within statistical uncertainties, which indicates no significant deviation from earlier constraints.

The best-fit value of $\varepsilon_{ee}-\varepsilon_{\mu\mu}$ is found to be $-0.59$, which matches with the value reported by the 3-year IceCube DeepCore analysis~\cite{IceCubeCollaboration:2021euf}. The p-value for the NSI fit in our analysis is 0.24. The observed best-fit value corresponds to a $\Delta\chi^2_{\rm SI-NSI}$ of 0.25 relative to the standard interaction case, which indicates no significant preference for the NSI hypothesis. The expected sensitivity shown by the dashed black curve is calculated using the simulated data as truth, where the best-fit values from the data fitting are used. The observed result is consistent with the expected sensitivity. The sensitivity to $\varepsilon_{ee}-\varepsilon_{\mu\mu}$ is limited, as it mainly affects the $\nu_e$ appearance channel, which has relatively low statistics in the present data sample. No limits at 90\% CL are placed on $\varepsilon_{ee}-\varepsilon_{\mu\mu}$ due to the limited sensitivity. However, values outside the interval $[-1.43\,,-1.11]\cup[-0.8\,,1.08]\cup[1.20\,,1.37]$ are excluded at 68.3\% confidence level. The green and yellow bands represent the 68.3\% and 90\% ranges for the expected $\Delta\chi^2$, respectively, where fit is performed over the statistically fluctuated pseudoexperiments. The observed $\Delta\chi^2$ profiles lie within these bands, which indicates consistency with statistical expectations.

\section{Global comparison}
\label{sec:Global_Comparison}

\begin{table}[tp!]
	\centering
	\renewcommand{\arraystretch}{1.2}
	\begin{tabular}{l@{\hskip 15pt}c}
		\hline
		\hline
		Parameters                                                                   & Bounds at 90\% CL     \\
		\hline
		$|\varepsilon_{e\mu}|$                                                      & $\leq 0.12$         \\
		$\varepsilon_{e\mu}$                                                        & $[-0.11,\,0.12]$     \\
		$|\varepsilon_{e\tau}|$                                                     & $\leq 0.18$          \\
		$\varepsilon_{e\tau}$                                                       & $[-0.11,\,0.18]$     \\
		$\varepsilon_{ee}-\varepsilon_{\mu\mu}$                                     & -                    \\
		$|\varepsilon_{\mu\tau}|$~\cite{Krishnamoorthi:2025efw}                     & $\leq 0.018$         \\
		$\varepsilon_{\mu\tau}$~\cite{Krishnamoorthi:2025efw}                       & $[-0.0094,\,0.0079]$ \\
		$\varepsilon_{\tau\tau}-\varepsilon_{\mu\mu}$~\cite{Krishnamoorthi:2025efw} & $[-0.031,\, 0.029]$  \\
		\hline
		\hline
	\end{tabular}
	\caption{The observed 90\% CL allowed ranges for the real-valued flavor-violating NSI parameters $\varepsilon_{e\mu}$ and $\varepsilon_{e\tau}$, as well as the nonuniversal NSI parameter $\varepsilon_{ee}-\varepsilon_{\mu\mu}$, obtained in this analysis using the 8-year golden event sample of IceCube DeepCore. For completeness, we have also included the bounds on $\varepsilon_{\mu\tau}$ and $\varepsilon_{\tau\tau}-\varepsilon_{\mu\mu}$ from our previous analysis~\cite{Krishnamoorthi:2025efw}.}
	\label{tab:real_bound}
\end{table}

\begin{figure*}[htp!]
	\centering
	\includegraphics[width=\textwidth]{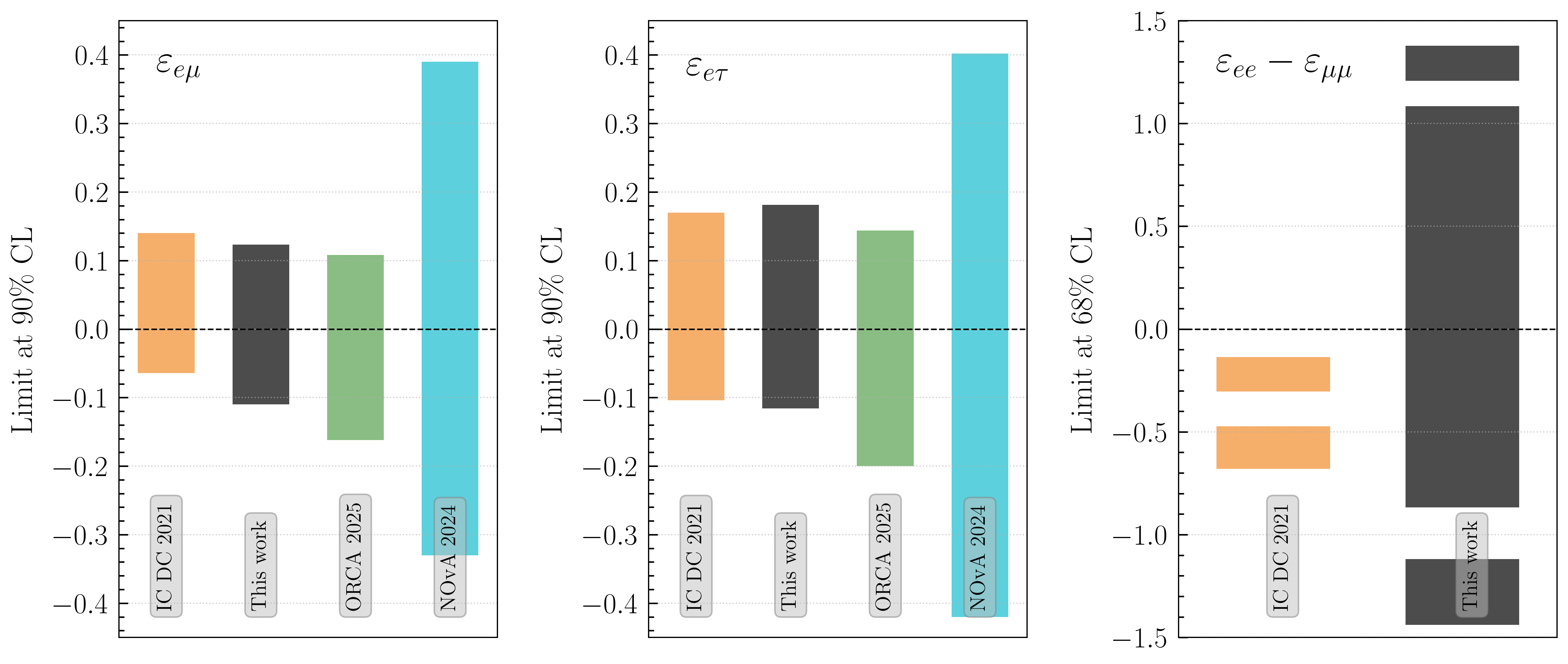}
	\caption{Constraints on the real-valued NSI parameters $\varepsilon_{e\mu}$ and $\varepsilon_{e\tau}$ at 90\% CL, and $\varepsilon_{ee}-\varepsilon_{\mu\mu}$ at 68\% CL from this analysis using the 8-year golden event sample of IceCube DeepCore. The NSI limits from IceCube DeepCore (2021)~\cite{IceCubeCollaboration:2021euf}, KM3NeT/ORCA (2025)~\cite{KM3NeT:2024pte}, and NOvA (2024)~\cite{NOvA:2024lti} are also include for comparison. For $\varepsilon_{e\mu}$ and $\varepsilon_{e\tau}$, the limits are evaluated by fixing the complex phase to $0$ (for positive values) and $\pi$ (for negative values). For consistency, bounds from some experiments have been converted to the NSI convention used in our analysis.}
	\label{fig:nsi_comparison}
\end{figure*}

The constraints obtained in this analysis for the NSI parameters $\varepsilon_{e\mu}$, $\varepsilon_{e\tau}$, and $\varepsilon_{ee}-\varepsilon_{\mu\mu}$ using the 8-year golden event sample of IceCube DeepCore  are summarized in the Tab.~\ref{tab:real_bound}, where the bounds are reported at 90\% confidence level. For the parameter $\varepsilon_{ee}-\varepsilon_{\mu\mu}$, we don't place any meaningful constraint due to the limited sensitivity. The bounds on the NSI parameters $\varepsilon_{\mu\tau}$ and $\varepsilon_{\tau\tau}-\varepsilon_{\mu\mu}$ are also included from our previous work~\cite{Krishnamoorthi:2025efw} for the completeness. Figure~\ref{fig:nsi_comparison} shows the summary of constraints on the real-valued NSI parameters obtained in this analysis, and comparison is done with the NSI limits from atmospheric and accelerator-based neutrino experiments. Overall, the bounds from this analysis are comparable and consistent with the previous IceCube DeepCore results. Note that our analysis uses the high-purity $\nu_\mu$ CC data sample, which is mainly contributed by $\nu_\mu \rightarrow \nu_\mu$ disappearance channel that provides the NSI bounds free from $\delta_{CP}$ degeneracy. Our analysis provides complementary bounds to those from the long-baseline experiments.


\section{Conclusions}
\label{sec:conclusion}

In this paper, we have performed a comprehensive study to constrain the NC-NSI parameters $\varepsilon_{e\mu}$, $\varepsilon_{e\tau}$, and $\varepsilon_{ee}-\varepsilon_{\mu\mu}$ using the publicly available atmospheric neutrino data from IceCube DeepCore with a livetime of 7.5 years. These NSI parameters are probed one at a time while considering other NSI parameters to be zero. We show that the data prefers the standard interaction hypothesis without any major deviations and find no evidence for NSI. The constraints on these parameters are reported  at 90\% CL and compared with those from atmospheric neutrino experiments such as IceCube DeepCore and KM3NeT/ORCA, as well as accelerator-based long-baseline neutrino oscillation experiment NO$\nu$A. This is a merit of a high-purity $\nu_{\mu}$ CC sample that, in spite of having relatively low statistics, the bounds from this work are comparable and consistent with bounds from the above-mentioned experiments. Another advantage of the high-purity $\nu_{\mu}$ CC sample is that the events are mainly contributed by the $\nu_\mu \rightarrow \nu_\mu$ disappearance channel, which provides the NSI constraints free from $\delta_{CP}$-degeneracy. Therefore, these measurements are complementary to those from the long-baseline neutrino oscillation experiments, where the appearance channel has a significant dependence on $\delta_{CP}$. Our study shows that the atmospheric neutrino data can play a crucial role in constraining these NC-NSI parameters. The upcoming good quality data from the atmospheric neutrino experiments like KM3NeT/ORCA, IceCube Upgrade, Hyper-K, and DUNE would enrich this field and provide more information in probing these NSI parameters.


\section*{Acknowledgements}\label{sec:acknowledgement}
S.K.A., J.K., and A.K. acknowledge support from the Department of Atomic Energy (DAE), Govt. of India, and the Swarnajayanti Fellowship (Sanction Order No. DST/SJF/PSA-05/2019-20) provided by the Department of Science and Technology (DST), Govt. of India, and the Research Grant (Sanction Order No. SB/SJF/2020-21/21) provided by the Anusandhan National Research Foundation (ANRF), Govt. of India, under the Swarnajayanti Fellowship. The numerical simulations are performed using the Dell PowerEdge R660 Server and the “SAMKHYA: High-Performance Computing Facility” at the Institute of Physics,  Bhubaneswar, India.

\section*{Data availability} 
In this work, we analyze the publicly available data from the IceCube Collaboration~\cite{DVN_B4RITM_2025}. Constraints on the NSI parameters obtained in this study are available from the authors in the form of digitized files upon reasonable request. 
\appendix
\section{SYSTEMATIC PARAMETERS}
\label{app:systematic_params}

In the present work, we incorporate the systematic uncertainties related to the detector response, Honda flux for atmospheric neutrinos~\cite{Honda:2015fha}, neutrino interaction cross section, normalization, and neutrino oscillations following the treatment described in Ref.~\cite{IceCubeCollaboration:2023wtb}. Table~\ref{tab:systematic_params} shows the list of the systematic uncertainty parameters that are considered as free nuisance parameters during the fit in the present analysis. The fifth and sixth columns show the nominal value and the corresponding $1\sigma$ prior (if available) for each parameter. We add a pull penalty term to the $\chi^2_{\rm mod}$ for the nuisance parameter where a prior is available. If no prior is available then the parameter is assumed to have a uniform distribution without any pull penalty. The second, third, and fourth columns show the best-fit values for these nuisance parameters that are obtained after fitting the observed data with the SI + NSI hypothesis having parameter $\varepsilon_{e\mu}$, $\varepsilon_{e\tau}$, and $\varepsilon_{ee}-\varepsilon_{\mu\mu}$, respectively, considering one at a time.


\section{DATA-MC COMPARISON}
\label{app:data_mc}

\setcounter{figure}{0}
\renewcommand{\thefigure}{B\arabic{figure}}
\renewcommand{\theHfigure}{B\arabic{figure}}
\begin{figure}[htp!]
	\includegraphics[width=\linewidth]{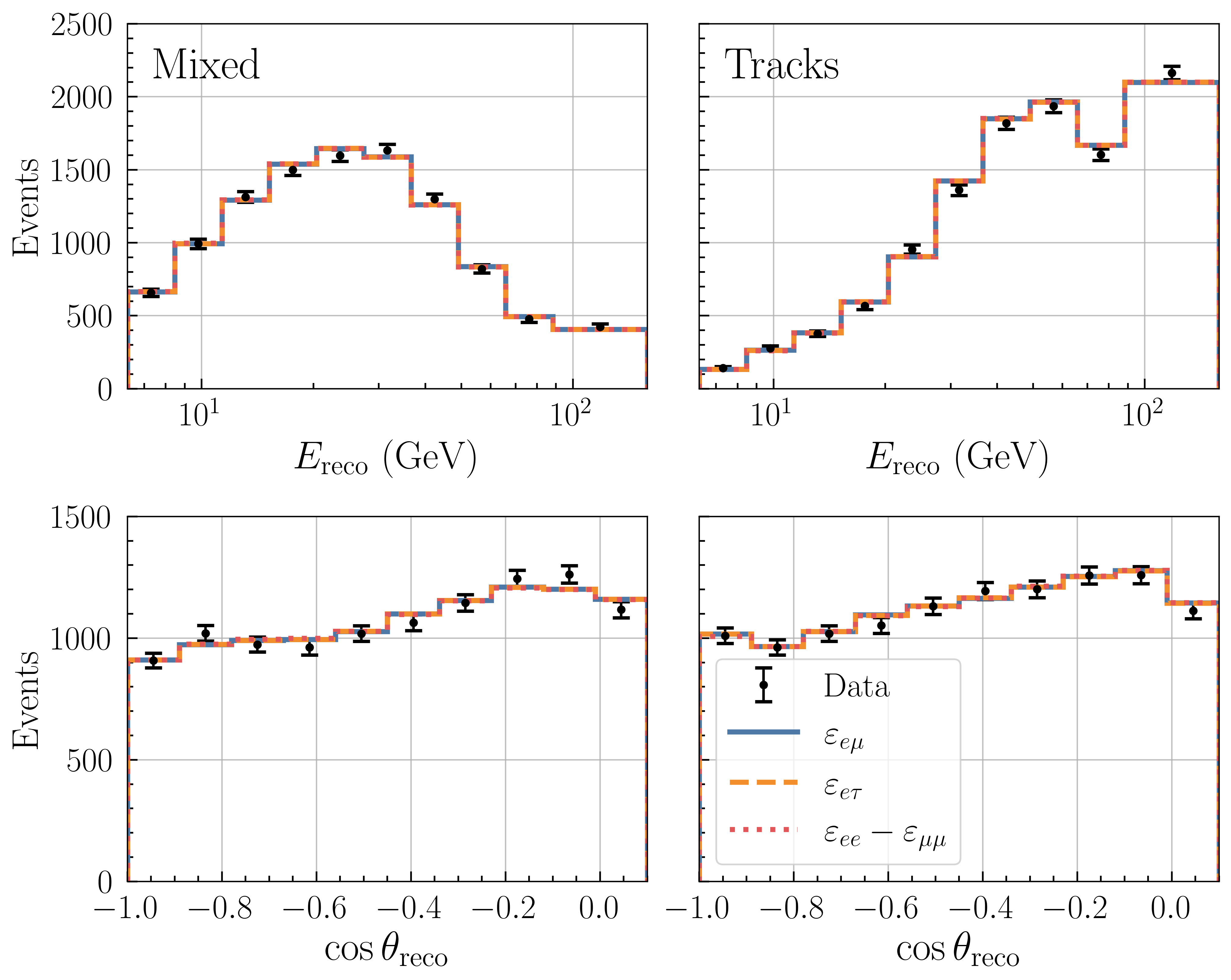}
	\caption{1D projections of the observed data and MC predictions with best-fit values as functions of the reconstructed energy (top panels) and cosine of zenith angle (bottom panels) for the 8-year golden event sample of IceCube DeepCore. The left and right panels present the mixed and tracklike events, respectively. The black dots with error bars denote observed data points. The solid blue, dashed orange, and dotted red histograms correspond to the best-fit MC predictions for the three SI + NSI hypotheses with $\varepsilon_{e\mu}$, $\varepsilon_{e\tau}$, and $\varepsilon_{ee}-\varepsilon_{\mu\mu}$, respectively, considering one at a time.}
	\label{fig:data_mc_comparison}
\end{figure}

In this section, we compare the observed data in the 8-year golden event sample of IceCube DeepCore and the MC predictions using best-fit values considering the SI + NSI hypotheses in the fit. Figure~\ref{fig:data_mc_comparison} shows the 1D projections of the observed data and MC predictions with the best-fit values as functions of reconstructed energy (top panels) and cosine of zenith angle (bottom panels). The left and right panels show the events with mixed and tracklike topologies, respectively. The black dots with error bars denote the observed data points. The solid blue, dashed orange, and dotted red histograms show the MC predictions using the best-fit values corresponding to the SI + NSI hypothesis with $\varepsilon_{e\mu}$, $\varepsilon_{e\tau}$, and $\varepsilon_{ee}-\varepsilon_{\mu\mu}$, respectively, considering one NSI parameter at a time. The MC predictions with the best-fit values are in good agreement with the observed data across all bins.

\onecolumngrid

\setcounter{table}{0}
\renewcommand{\thetable}{A\arabic{table}}
\renewcommand{\theHtable}{A\arabic{table}}
\begin{table}[ht]
	\centering
	\renewcommand{\arraystretch}{1.3}
	\begin{tabular}{l@{\hskip 4pt}c@{\hskip 4pt}c@{\hskip 3pt}c@{\hskip 3pt}c@{\hskip 5pt}c}
		\hline
		\hline
		\textbf{Parameters}                  & \multicolumn{3}{c}{\textbf{Best-fit values}}     & \textbf{Nominal values} & \textbf{\quad Priors ($1\sigma$)}                                 \\
		& $\varepsilon_{e\mu}$ & $\varepsilon_{e\tau}$ & $\varepsilon_{ee}-\varepsilon_{\mu\mu}$   &                 &             \\
		\hline
		\multicolumn{6}{@{}l}{\textbf{Detector:}}                                                                                                                             \\
		DOM efficiency                                 & 1.062                  & 1.064       & 1.053                & 1.0             & $\pm~$0.1   \\
		Ice absorption                                  & 0.973                  & 0.973      & 0.970                & 1.0             & Unconstrained           \\
		Ice scattering                                  & 0.988                  & 0.989      & 0.990                & 1.05            & Unconstrained           \\
		Relative eff. $p_0$                        & $-\,0.264$             & $-\,0.264$      & $-\,0.232$           & 0.10            & Unconstrained           \\
		Relative eff. $p_1$                        & $-\,0.042$             & $-\,0.044$      & $-\,0.043$           & $-\,0.05$       & Unconstrained           \\
		\hline
		\multicolumn{6}{@{}l}{\textbf{Atmospheric neutrino flux:}}                                                                                                            \\
		$\Delta \gamma_\nu$                             & 0.063                  & 0.063      & 0.063                & 0.0             & $\pm\,$0.1  \\
		$\Delta \pi^+ \text{ yields [A-F]}$             & 0.059                  & 0.058      & 0.056                & 0.0             & $\pm\,$0.3  \\
		$\Delta \pi^+ \text{ yields G}$            & $-\,0.061$             & $-\,0.059$      & $-\,0.066$           & 0.0             & $\pm\,$0.3 \\
		$\Delta \pi^+ \text{ yields H}$            & $-\,0.021$             & $-\,0.019$      & $-\,0.026$           & 0.0             & $\pm\,$0.15 \\
		$\Delta K^+ \text{ yields W}$                   & 0.089                  & 0.088      & 0.091                & 0.0             & $\pm\,$0.4  \\
		$\Delta K^+ \text{ yields Y}$                   & 0.115                  & 0.109      & 0.125                & 0.0             & $\pm\,$0.3  \\
		$\Delta K^- \text{ yields W}$              & $-\,0.007$             & $-\,0.010$      & $-\,0.004$           & 0.0             & $\pm\,$0.4  \\
		\hline
		\multicolumn{6}{@{}l}{\textbf{Neutrino interaction cross section:}}                                                                                                   \\
		$M_A^{\text{CCQE}}$ (in $\sigma$)               & 0.065                  & 0.088      & 0.067                & 0.0             & $\pm\,$1.0  \\
		$M_A^{\text{CCRES}}$ (in $\sigma$)              & 0.619                  & 0.640      & 0.654                & 0.0             & $\pm\,$1.0  \\
		DIS CSMS                                        & 0.055                  & 0.065      & 0.130                & 0.0             & $\pm\,$1.0  \\
		$\sigma_{\rm NC}/\sigma_{\rm CC} $              & 1.123                  & 1.113      & 1.112                & 1.0             & $\pm\,$0.2  \\
		\hline
		\multicolumn{6}{@{}l}{\textbf{Normalization:}}                                                                                                                        \\
		$A_{\text{eff}}$ scale                          & 0.827                  & 0.825      & 0.831                & 1.0             & Unconstrained           \\
		\hline
		\multicolumn{6}{@{}l}{\textbf{Atmospheric muons:}}                                                                                                                    \\
		Atm. $\mu$ scale                                & 1.33                   & 1.321      & 1.371                & 1.0             & Unconstrained           \\
		\hline
		\multicolumn{6}{@{}l}{\textbf{Oscillations:}}                                                                                                                         \\
		$\theta_{23}$                          & 45.804$^\circ$         & 44.2027$^\circ$     & 45.622$^\circ$       & 45.573$^\circ$  & Unconstrained           \\
		$\Delta m^2_{31}$                     & 0.002511 eV$^2$        & 0.002573 eV$^2$      & 0.002521 eV$^2$      & 0.002484 eV$^2$ & Unconstrained           \\
		\hline
		\hline
		\end{tabular}
		
		\caption{The table presents the systematic uncertainty parameters that are considered as nuisance parameters in the present analysis. The fifth and sixth columns show the nominal value and the associated $1\sigma$ prior (if available), respectively, for each parameter. The table also presents the best-fit values for these nuisance parameters that are obtained after fitting the data with the SI + NSI hypothesis having parameter,  $\varepsilon_{e\mu}$ (second column), $\varepsilon_{e\tau}$ (third column), and $\varepsilon_{ee}-\varepsilon_{\mu\mu}$ (fourth column), considering one at a time.}
		\label{tab:systematic_params}
	\end{table}
	\twocolumngrid


%

\end{document}